\documentclass[reprint,aps,prx,letterpaper,longbibliography]{revtex4-2}
\usepackage{amssymb}
\usepackage{amsmath}
\usepackage{graphicx}
\usepackage{footnote}
\usepackage{subcaption}
\usepackage[usenames,dvipsnames,svgnames,table]{xcolor}
\usepackage[T1]{fontenc}
\usepackage[
pdfauthor={Victoria Zhang, Paul D. Nation},
pdftitle={Characterizing quantum processors using discrete time crystals},
bookmarks=true,colorlinks=true,linkcolor=OrangeRed,urlcolor=RoyalBlue,citecolor=DarkOrchid]{hyperref}

\definecolor{lightgray}{gray}{0.95}
\let\oldtabular\tabular
\let\endoldtabular\endtabular
\renewenvironment{tabular}{\rowcolors{2}{white}{lightgray}\oldtabular}{\endoldtabular}

\begin{document}

\title{Characterizing quantum processors using discrete time crystals}

\author{Victoria Zhang}
\thanks{Present address: Harvard University, Cambridge, MA 02138, USA}
\author{Paul D. Nation}
\email[E-mail: ]{paul.nation@ibm.com}
\affiliation{IBM Quantum, Yorktown Heights, NY 10598, USA}

\begin{abstract}
We present a method for characterizing the performance of noisy quantum processors using discrete time crystals.  Deviations from ideal persistent oscillatory behavior give rise to numerical scores by which relative quantum processor capabilities can be measured.  We construct small sets of qubit layouts that cover the full topology of a target system, and execute our metric over these sets on a wide range of IBM Quantum processors.  We show that there is a large variability in scores, not only across multiple processors, but between different circuit layouts over individual devices.  The stability of results is also examined.  Our results suggest that capturing the true performance characteristics of a quantum system requires interrogation over the full device, rather than isolated subgraphs.  Moreover, the disagreement between our results and other metrics indicates that benchmarks computed infrequently are not indicative of the real-world performance of a quantum processor.  This method is platform agnostic, simple to implement, and scalable to any number of qubits forming a linear-chain, while simultaneously allowing for identifying ill-performing regions of a device at the individual qubit level.   
\end{abstract}
\date{\today}

\maketitle

\section{Introduction}\label{sec:intro}
Computing the performance characteristics of quantum processors has long been a mainstay of quantum computing research \cite{magesan:2012b,magesan:2012,magesan:2011,knill:2008}.  More recently, as quantum devices have reached qubit counts beyond these more rigorous methods, focus has moved to holistic benchmarks \cite{proctor:2022b, cross:2019} and suites of test-circuits \cite{proctor:2022,tomesh:2022,lubinsky:2021,kohout:2020,li:2020} to validate progress.  With the public availability of multiple quantum processor technologies, these ``application benchmarks'' allow for performance comparisons across platforms that take into account the strengths and weaknesses of each approach.

Here we propose a new characterization method based on recent demonstrations of discrete time crystals (DTC) \cite{frey:2022,mi:2022}.  DTC has several properties that make it suitable for characterizing quantum hardware.  First, the non-equilibrium DTC state of an ideal periodically driven many-body system exhibits stable oscillations of the single-qubit polarization $\langle Z_{i}(t)\rangle$, deviations from which indicate imperfections in the  system (noise), and provide a natural means by which to characterize processor performance.  Second, unlike benchmarks such as Quantum Volume (QV) \cite{cross:2019} that effectively twirl away coherent errors \cite{wallman:2016,knill:2004}, DTC evolution is generated by periodic applications of a unitary that enhances coherent errors in the system, and is thus sensitive to noise sources that benchmarks with random structure might otherwise miss.  Moreover, in contrast to many Hamiltonian simulation test-circuits that need some level of Trotterization specified, DTC evolution is exact, simplifying the analysis.

In this work we show that DTC allows for in-depth characterization of quantum processor performance.  Unlike most quantum benchmarks and test suites, where device performance is gauged on only a subset of qubits, we are interested in both the local and device-wide performance of a quantum processor.  While full system characterization requires running circuits on every compatible subgraph of the full device topology; a task that scales unfavorably as device connectivity and number of qubits grows, here we take advantage of the fact that computing a set of physical qubit layouts (subgraphs) corresponding to a given input circuit is fast relative to other compilation timescales \cite{nation:2022}, and use a minimal, or close to minimal, set of layouts that extends over the full topology of the device as a proxy for complete system characterization.  

We execute our DTC benchmark on the majority of IBM Quantum systems, and compare the performance across these processors.  In particular, we find that there are large performance variations over quantum systems with numbers of qubits large enough to support multiple disjoint layouts.  Additionally we see that other stated metric values attached to each processor are, in general, not good predictors of DTC circuit performance.   The locality of noise in systems with many-body localization \cite{ippoliti:2021,serbyn:2013,bardarson:2012} allows variations across a device to be traced to individual qubits that perform poorly, indicating a strong local noise environment; qubits act as regional noise barometers.  Finally, we also look at stability of our benchmark over time and find that, while there are large variations in the absolute performance scores for a given qubit layout, including within a single calibration cycle, sections of the chip with the highest relative reported values tend to remain optimal over time.  Our method is platform agnostic and simple to implement, providing a faithful metric by which relative quantum processor performance can be gauged and tracked over time.

The paper is organized as follows.  First, in Sec.~(\ref{sec:method}) we describe our method of DTC characterization focusing on circuit creation, performance metric definition, and detailing the procedure for layout construction used for full-device performance evaluation.  Section~(\ref{sec:results}) looks at the results of running $5$ and $20+$ qubit DTC performance evaluations over IBM Quantum systems, including a discussion of the stability of results over time.  Finally, we summarize the results in Sec.~(\ref{sec:conclusion}), discuss how to interpret these findings, and point to possible future research directions utilizing the methods presented here.  Appendix~(\ref{app:noise}) includes a discussion of noise localization in systems such as DTC, including a simulation highlighting this localization for the circuits used in this work.

\section{Methodology}\label{sec:method}
Our starting point is the unitary operator for a DTC corresponding to a linear-nearest neighbor Hamiltonian with discrete time-translation symmetry, Eq.~(1) from Ref.~\cite{mi:2022}
\begin{equation}\label{eq:main}
\hat{U} = e^{-\frac{i}{2}\sum_{j}h_{j}\hat{Z}_{j}}~e^{-\frac{i}{4}\sum_{j}\phi_{j}\hat{Z}_{j}\hat{Z}_{j+1}}~e^{-i\frac{\pi g}{2}\sum_{j}\hat{X}_{j}},
\end{equation}
where $g$ controls the angle of rotation for all qubits about the x-axis, and $\phi_{j}$ and $h_{j}$ are phases drawn at random from a uniform distribution whose average value is dependent on the selected value of $g$ \cite{khemani:2016}.  While the use of random parameters is required to stabilize the many-body localization needed for observing DTC \cite{ippoliti:2021}, the ability to create random instances of circuits is also a desired property for characterizing quantum systems.

Our characterization circuits are formed by generating a specific realization of Eq.~(\ref{eq:main}), $\hat{\mathcal{U}}$, specified by the number of qubits $q$, a given $g\sim 1$, and a corresponding random set of phases set by a random number generator seeded with a given integer.  This unitary is then used to generate a family of circuits, each with an increasing number of  $\hat{\mathcal{U}}$ applied to a specific initial state $|\psi(0)\rangle$: $\{ \hat{\mathcal{U}}^{n}|\psi(0)\rangle ~|~ n \in [0, N_{\rm max}]\}$, where $N_{\rm max}$ is the maximum number of discrete cycles we wish to observe.  While we are free to pick any initial state, here we select the ground state of the system $|\psi(0)\rangle = |0^{\otimes q}\rangle$, as any deviations from ideal behavior when no $\hat{\mathcal{U}}$ are applied can then be directly equated to state preparation and measurement (SPAM) errors in the system.

\begin{figure}[b]
\includegraphics[width=7.5cm]{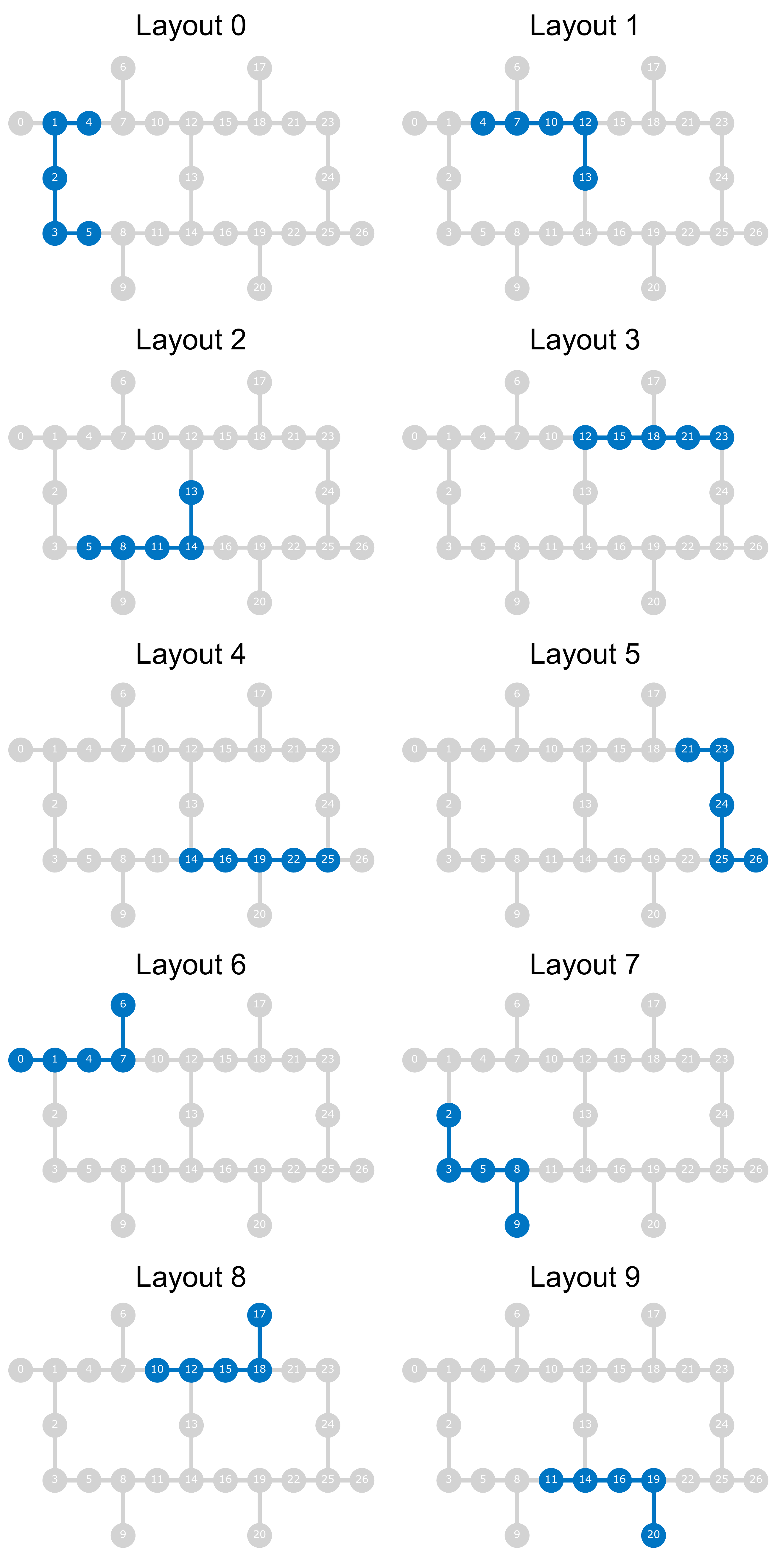}
\caption{Layouts corresponding to the 5-qubit covering set used on IBM Quantum 27-qubit Falcon processors in this work.}
\label{fig:falcon_covering}
\end{figure}

We are interested in the \textit{cycle amplitude} $A_{j}(n)$ of qubit $j$, defined to be the absolute difference in polarization values between the $n_{\rm th}$ and $n_{\rm th}+1$ cycles: $A_{j}(n) =  \left| \langle Z_{j}(n+1)\rangle-\langle Z_{j}(n)\rangle\right|$.  From simple depolarizing noise model simulations it is expected that the cycle amplitude decays exponentially, suggesting that we can define a cycle visibility threshold of $2/e$, below which the cycle amplitude $A_{j}(n)$ is considered no longer visible.  The maximum number of cycles for which a given  $A_{j}(n)$ is above this visibility threshold is defined to be the number of \textit{visible cycles}, $N_{\rm vis}$, and is the quantity of interest in our characterization. 

Noise localization in DTC allows us to connect the number of visible cycles to the local errors acting on a qubit.  This is easiest seen in the small-error limit of the depolarizing model \cite{ippoliti:2021}, and ignoring single-qubit gate errors that are over an order of magnitude smaller than those of two-qubit gates on IBM Quantum systems. This leads to an approximate expression for the number of visible cycles on the $i$-th qubit $N^{(i)}_{\rm vis}$:
\begin{equation}\label{eq:relation}
N^{(i)}_{\rm vis} \simeq \frac{15}{16}\frac{g(\Gamma_{i})}{\left(\epsilon_{i}^{\rm left} + \epsilon_{i}^{\rm right}\right)},
\end{equation}
where $\epsilon_{i}^{\rm left}$ and $\epsilon_{i}^{\rm right}$ are the two-qubit entangling gate errors to the left and right of the $i$-th qubit in the chain, respectively, and $g(\Gamma_{i}) \lesssim 1$ relates the qubit readout-error, $\Gamma_{i}$, in the weak-error limit to the corresponding reduction in the cycle-amplitude.  Note that, while Eq.~(\ref{eq:relation}) makes clear the functional relationship between the number of cycles and the error rates acting on a qubit, in practice this expression need not be valid for all qubits in a device.  Indeed, we will encounter examples of both strong readout and gate errors in Sec.~(\ref{eq:relation}).

For comparing results over multiple qubits, we use the mean number of visible cycles over the set of qubits.  This value does not require classical simulation to evaluate, allowing our metric to scale beyond $\mathcal{O}(10)$ qubits.  Note that we could also derive a metric based on values for the decay rate in the cycle amplitude curves found via curve fitting.  However in practice, deviations from ideal exponential decay can lead to difficulties in fitting, and curve fitting requires a large number of circuit evaluations.  In contrast, although not done here, our metric based on cycle visibility allows for iterative execute of circuits only up to the point where visibility is lost; the cost of performing DTC characterization can be greatly reduced. 

Due to the wide variability in important system metrics seen in near-term quantum systems \cite{nation:2022}, reporting the number of visible cycles, or any other characterization values, over just a subset of qubits fails to capture many aspects of device performance.  While it is relatively fast to compute all possible subgraphs that are compatible with a given input circuit \cite{treinish:2021, vf2++}, the combinatorial explosion of possibilities renders execution on hardware unfeasible.  As a proxy for this full characterization we introduce a \textit{covering set} of physical qubit layouts, that corresponds to a reduced set of layouts that cover the full device entangling gate topology (coupling map).  In this work, this set is produced starting with an initial layout, e.g. as given by \texttt{rustworkx} \cite{rustworkx}, and adding subsequent layouts that maximize the relative compliment between the edges of the system topology already in the set and those in the considered layout.  This procedure is continued until all the edges of the device topology are in the set. This greedy algorithm does not guarantee that the covering set is a minimal, but in practice it often is, or very close to it.  An example covering set for 5-qubit DTC circuits over the topology for 27-qubit IBM Quantum Falcon systems is shown in Fig.~(\ref{fig:falcon_covering}).

\section{Results}\label{sec:results}

\subsection{5-qubit DTC characterization}\label{sec:results5}

\begin{figure}[t]
\includegraphics[width=7.5cm]{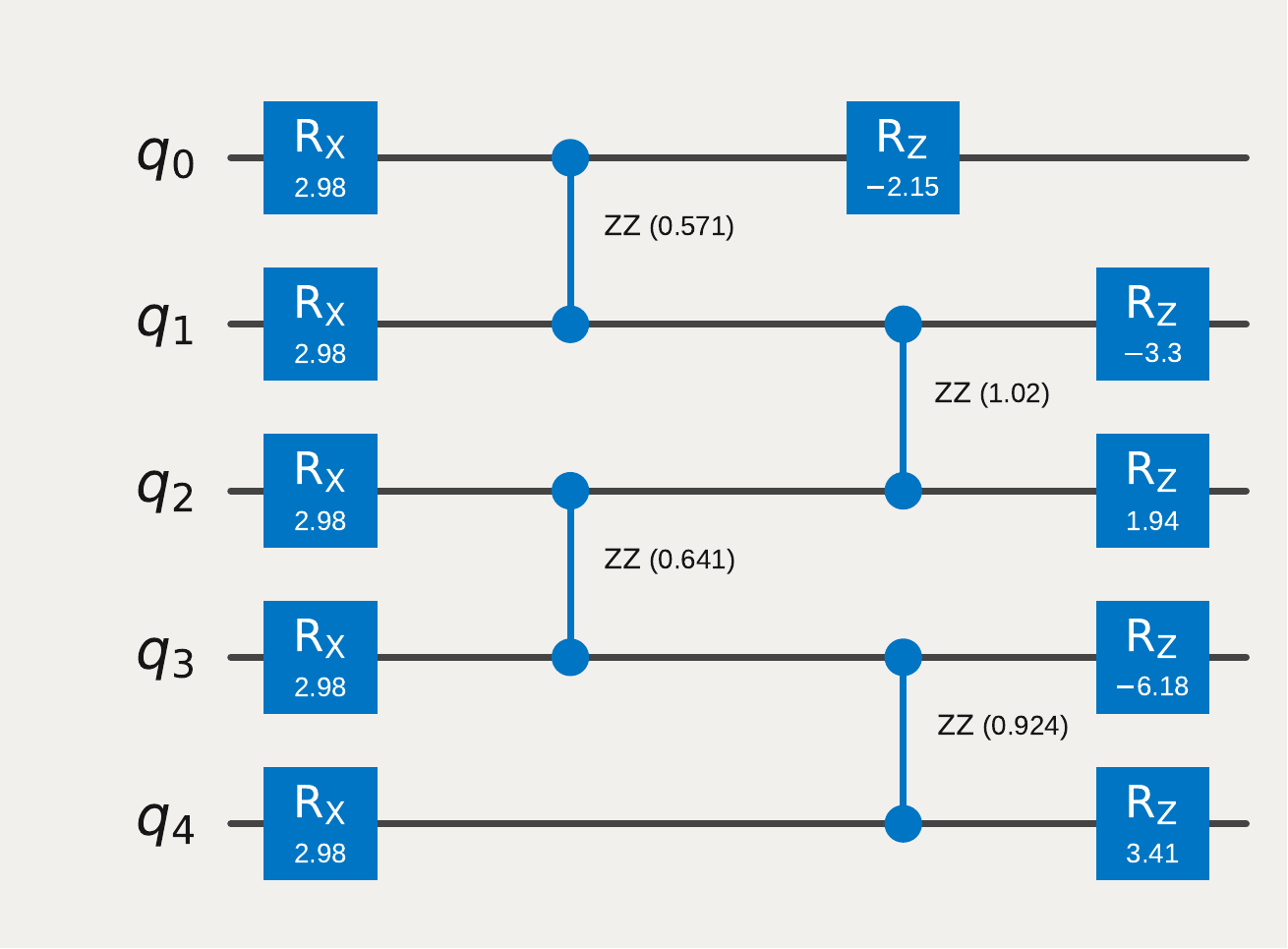}
\caption{The circuit for the 5-qubit DTC unitary operator used for characterization across systems.}
\label{fig:circ}
\end{figure}

To begin, we consider the number of visible cycles for a 5-qubit DTC characterization.  This allows for comparing our performance metric across the majority of IBM Quantum systems, and is also the same width as QV32 circuits, the minimal value for most processors, giving us a baseline for comparing our metric against the reported QV value for each system.  We generate DTC circuits \cite{crystalmark} out to $N_{\rm max}=80$ using Eq.~(\ref{eq:main}), by setting $g=0.95$, and $\rm RZZ$ gate parameters sampled uniformly from the interval $[\pi/8, 3\pi/8]$. Random $\rm RZ$ gate angles are sampled over the interval $[-\pi, \pi]$.  The corresponding circuit for $\hat{\mathcal{U}}$ generated using a random number generator with \texttt{seed=12345} is shown in Fig.~(\ref{fig:circ}).
Circuits are compiled using Qiskit \cite{qiskit} with \texttt{optimization\_level=3}, and mapped onto covering sets using \texttt{mapomatic} \cite{nation:2022,mapo}.  We use the Qiskit Runtime \texttt{Sampler} primitive for circuit execution, and jobs corresponding to individual layouts in a covering set are batched within a Runtime Session so that they execute sequentially on-chip; individual jobs are not subject to the fair-share scheduling priority.  We sample all circuits $10^{4}$ times for reasonable precision in the computed polarizations, and set \texttt{skip\_transpilation=True} as our circuits are already mapped to the target device.  Additionally, the default readout mitigation performed by the \texttt{Sampler} is disabled by setting \texttt{resilience\_level=0}.

\begin{figure}[b]
\includegraphics[width=\linewidth]{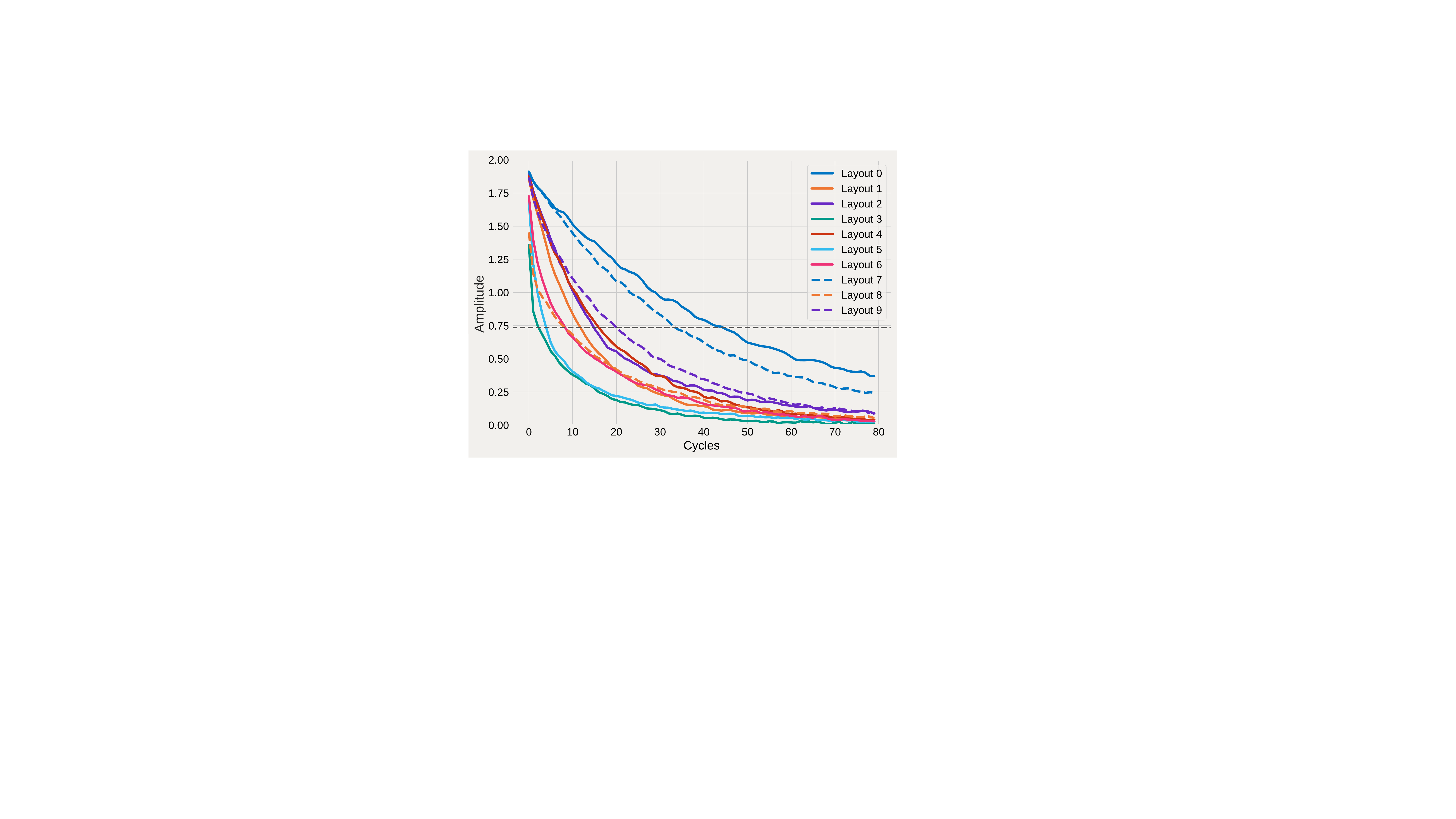}
\caption{Mean amplitude of layouts in a covering set, Fig.~(\ref{fig:falcon_covering}), as a function of the number of DTC cycles on the 27-qubit IBM Quantum Auckland system.  Dashed horizontal line shows the $2/e$ threshold for visibility.}
\label{fig:cycles-auckland}
\end{figure}

\begin{figure}[b]
\includegraphics[width=\linewidth]{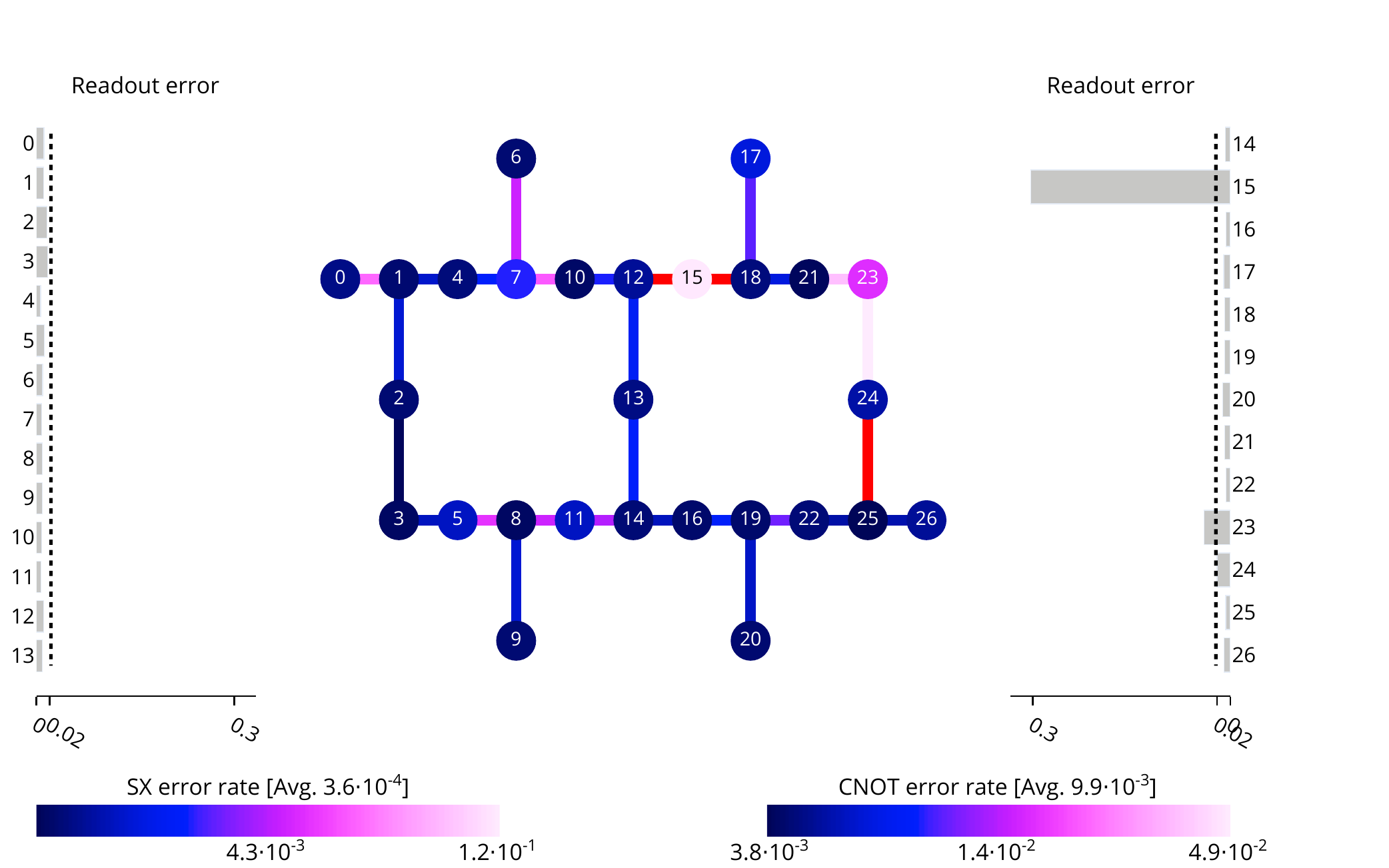}
\caption{Error map for the 27Q IBM Quantum Auckland system generated from calibration data taken on October 31, 2022.  Edges of the graph colored red indicate CNOT gates that failed the calibration procedure; nominally a signature of gates with outsized error rates.}
\label{fig:error_map}
\end{figure}

Figure~(\ref{fig:cycles-auckland}) shows the mean cycle visibility for the 10 layout covering set from Fig.~(\ref{fig:falcon_covering}) using the unitary $\hat{\mathcal{U}}$ in Fig.~(\ref{fig:circ}) on the 27-qubit IBM Quantum Auckland system.
We see that there is a marked difference in the average cycle amplitude decay rates across the layouts, highlighting the variability in performance across the system; something that would be missed if we did not probe the full topology.  This in turn gives rise to visible cycles ranging from 45 cycles (layout 0) down to 3 (layout 3), yielding a system wide average visibility of 16.7 cycles.  Layouts that deviate from the ideal value ($2$) at zero cycles, e.g. layouts 3 and 8 in Fig.~(\ref{fig:cycles-auckland}),  where no $\hat{\mathcal{U}}$ has been applied indicate that the qubits are not in the ground state as expected, and that SPAM errors are non-negligible over one or more qubits in the layouts, i.e. $g(\Gamma_{i}) <1$ in Eq.~(\ref{eq:relation}).  At the time of execution, Auckland calibration data, Fig.~(\ref{fig:error_map}), indicated that qubit 15 had a readout error rate of $\sim 30\%$. This qubit is in both layout 3 and 8, corroborating our model, and highlighting the sensitivity of DTC characterization on underlying device performance metrics.

\begin{figure}[t]
\includegraphics[width=\linewidth]{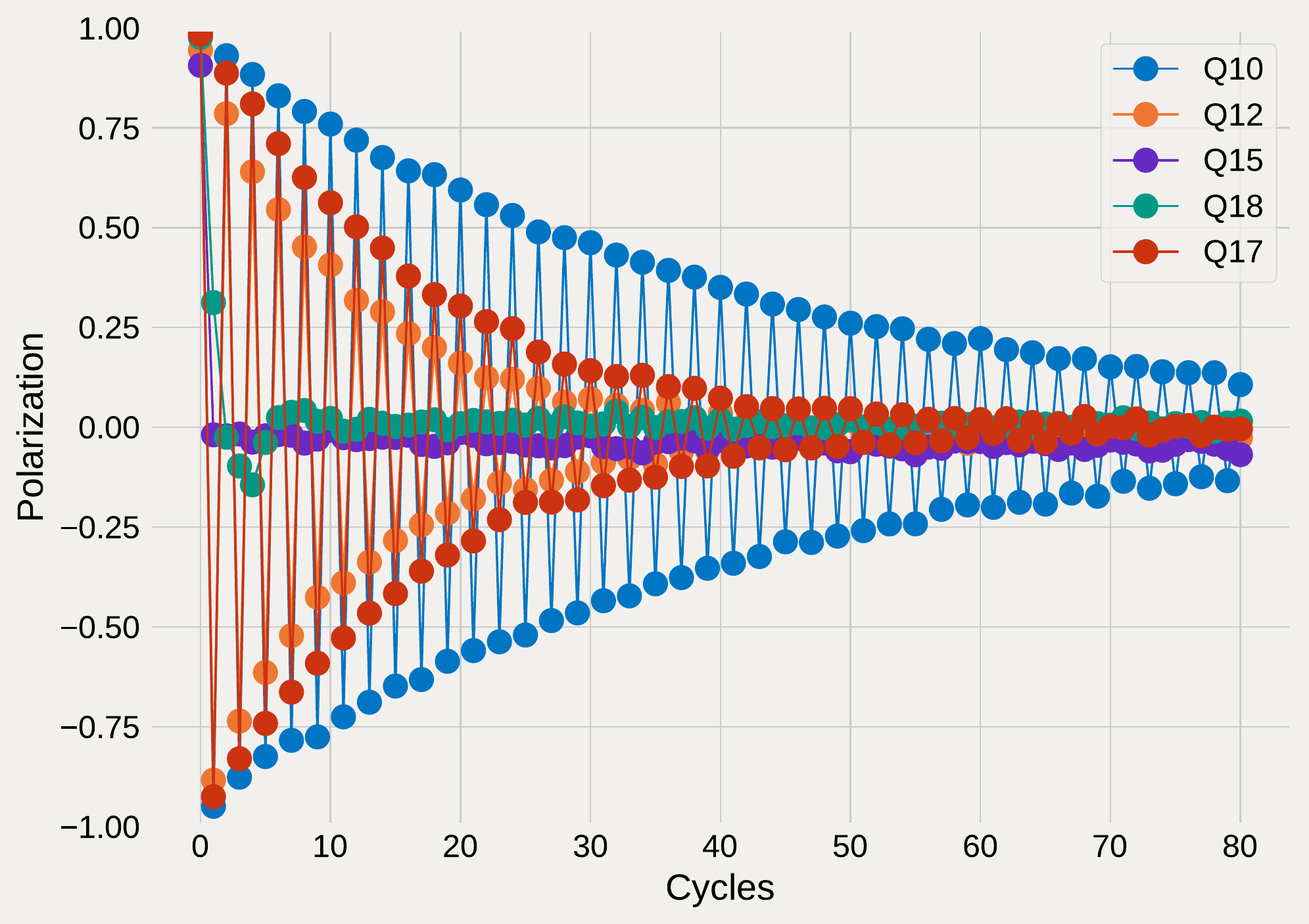}
\caption{Qubit polarization per cycle from Fig.~(\ref{fig:cycles-auckland}) for qubits $[10,12,15,18,17]$ forming layout 8.}
\label{fig:auckland-layout8}
\end{figure}

\begin{figure*}[t]
\includegraphics[width=5.5in]{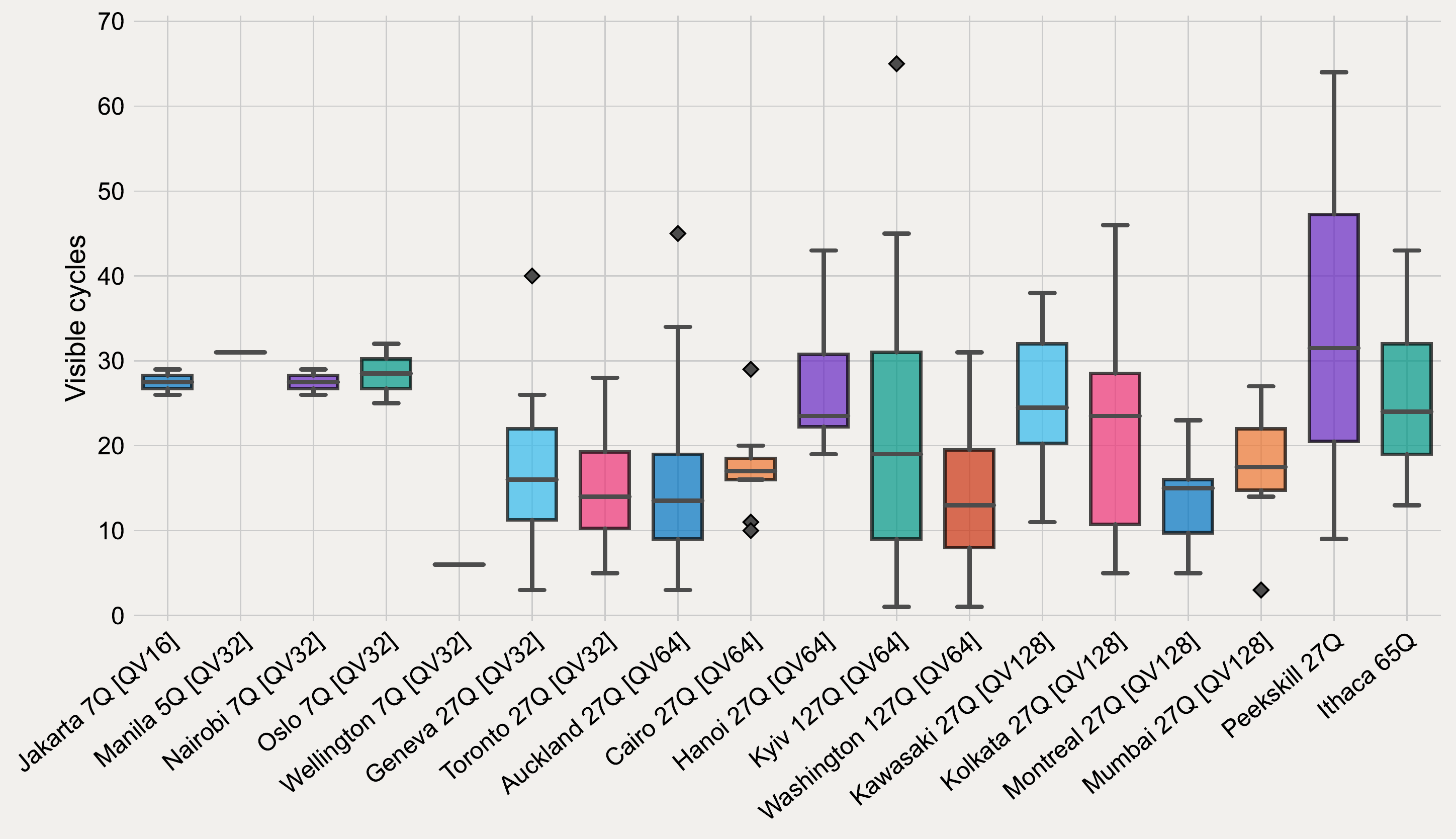}
\caption{Results of executing 5-qubit DTC characterization over a covering set for each quantum processor. The same covering set is used for processors with matching topologies.  Systems are ordered by increasing qubit count and reported QV values.  Processors with no reported QV values are presented last.}
\label{fig:5q_dtc}
\end{figure*}

In general, the poor performance of a given layout cannot be attributed to uniformly low-performing qubits.  Rather, it is nominally a small number of qubits within the given layout that swing the average lower.  As an example, in Fig.~(\ref{fig:auckland-layout8}) we plot the single-qubit polarization values for the qubits comprising layout 8 in Fig.~(\ref{fig:cycles-auckland}).  It is clear that two qubits, $\rm Q15$ and $\rm Q18$, fail to perform even a single meaningful oscillation, and thus greatly diminish the visibility of the set as a whole.  Once again, this agrees with the calibration data, Fig.~(\ref{fig:error_map}), where two out of the four CNOT gates in the layout failed calibration due to their large error rates.  Layout 3 also covers $\rm Q15$ and $\rm Q18$, see Fig.~(\ref{fig:falcon_covering}), and shows even lower cycle visibility, suggesting there are additional faulty qubits in that set.  That individual qubits can be corrupted without affecting the layout as a whole is a direct result of noise localization in many-body localized systems; qubits are primarily sensitive to only those noise sources that act directly on the qubit itself. A simulation of this noise localization is given in App.~(\ref{app:noise}).  From Eq.~(\ref{eq:relation}), our characterization based on average number of visible cycles can be approximately viewed as the average of the inverse gate errors acting on individual qubits in a layout, with each component weighted by a term proportional to the qubit readout error.  

Although much of the DTC characterization performance corresponds well to device calibration data, this is not universally the case.  For example, layout 4 in Fig.~(\ref{fig:falcon_covering}) covers a section of the Auckland system with error rates well below the device average [see Fig.~(\ref{fig:error_map})], yet the DTC characterization in Fig.~(\ref{fig:cycles-auckland}) reveals that this section of the chip performs poorly, failing to achieve double-digit numbers of visible cycles; system data derived from one- and two-qubit calibrations can fail to capture real-world performance.  This underscores the need for characterization methods like DTC in evaluating the quality of quantum systems.

Having examined DTC characterization of a single quantum system, we now turn to performance comparisons across multiple quantum processors.  We repeat the same characterization performed on the Auckland device across a wide range of compatible IBM Quantum systems with the results presented in Fig.~(\ref{fig:5q_dtc}).  The same covering set as Auckland was used for all 27-qubit machines, while a new covering set was generated for other unique system topologies.  Figure~(\ref{fig:5q_dtc}) shows that, like the earlier results on Auckland, there is a large variability in device performance on systems with 27+ qubits, where there is little overlap between the individual layouts within the covering sets.  Many processors include subgraphs with visibilities in the single-digits indicating that a wide range of systems have one or more qubits with strong local noise that can greatly shift the mean cycle visibility lower at the granularity considered here.  The systems with layouts supporting the largest visible cycles, the Kyiv and Peekskill systems, also have the largest fluctuations in scores indicating that, while sections of the devices perform well, there is a large degree of variability in terms of quality across the chips.  Users executing experiments on these devices would therefore experience vastly different performance depending on the qubit layout used in circuit compilation.  This emphasizes how device-wide characterization and qubit selection is critical for achieving high fidelity results on near-term quantum devices

Figure~(\ref{fig:5q_dtc}) also reveals that the measured system performance is, in general, uncorrelated to the reported QV values for the systems.  In particular out of the QV128 systems, only the Kolkata device contains a layout that bests lower QV systems in our tests, save for the single high-visibility outlier layout on the QV64 Kyiv system. In contrast, the QV128 Montreal and Mumbai systems are unable to match the performance of even the QV16 Jakarta system.  As a whole we see that the IBM Quantum Peekskill system, with no reported QV value, performs the best on average, with 4 layouts supporting 45+ visible cycles;  there are 5Q regions of the topology with relatively little noise.  We caution that the results presented in Fig.~(\ref{fig:5q_dtc}) do not equate to claims that our metric and QV benchmarking are at odds with each other.  Rather, the disparity can likely be attributed to the fact that the reported QV values for a given system are nominally computed once, usually at system bring-up, and therefore may not be indicative of device performance at the time of DTC testing.  Some reported QV values are more than one year old.  In addition, the QV score of a system is computed only over a very specific set of qubits deemed to be the best performing set at the time and thus QV, like other metrics reported on singular subgraphs, is not indicative of total system performance.

\begin{figure}[b]
\includegraphics[width=\linewidth]{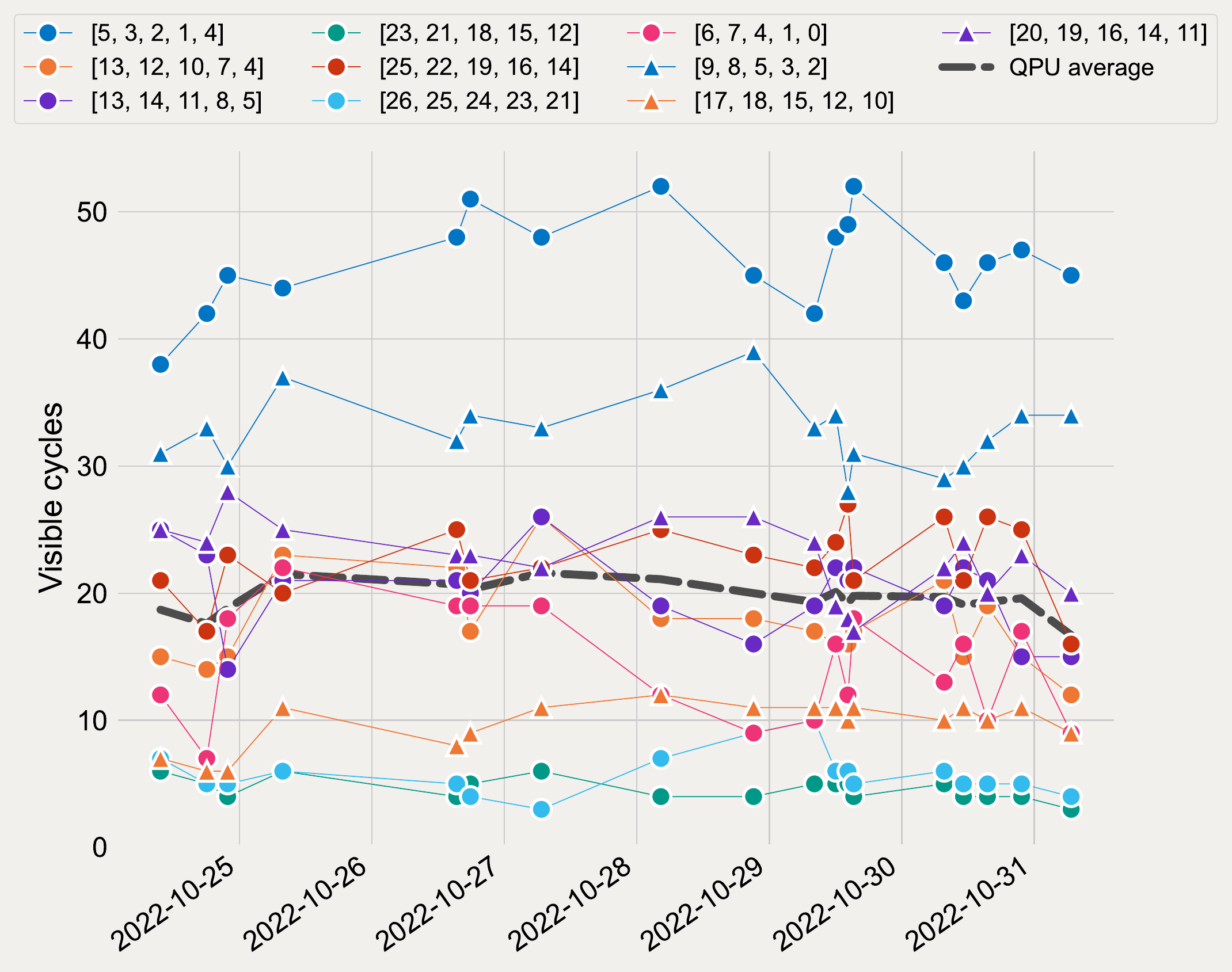}
\caption{Stability of DTC cycle visibility for the 10 layouts in the 5-qubit covering set, Fig.~(\ref{fig:falcon_covering}), on the IBM Quantum Auckland system.  Characterization was run repeatedly from October 24th to October 31st, 2022.  Variability between data points is due to lack of system availability, e.g. low scheduling priority, system reservations, calibration periods, etc.}
\label{fig:stability}
\end{figure}

While computing the QV of a system is difficult for users, especially QV64 and higher \cite{jurcevic:2021}, we can look at the stability of the system through the lens of DTC.  Figure~(\ref{fig:stability}) shows the number of visible cycles per layer in the covering set for IBM Quantum Auckland, measured over the course of a week.  It is clear that, even on the timescale of a single day, there are large swings, up to 11, in the number of visible cycles for layers capable of more than single-digit visibility.  Taken as a whole, the data for Auckland shows a visible cycle variability of $\sim \pm 20\%$ over the course of a week. However, while the absolute cycle visibility shows marked deviations, the most performant sets of qubits remain optimal despite of these fluctuations; the relative quality of sets of qubits is more consistent over time.   The variability seen in Fig.~(\ref{fig:stability}) gives credence to the idea that infrequently computed benchmarks like QV are not faithful predictors of current near-term system performance.  Taken together with Fig.~(\ref{fig:5q_dtc}), our results suggest that users looking to maximize the fidelity of their circuits could benefit from more frequent full-system characterization information. 

\subsection{20+ qubit DTC characterization}\label{sec:results20}
We now extend DTC characterization to layouts with 20+ qubits.  With our metric equating to average of the inverse noise strength experienced by qubits within a given layout, we expect a reduction in the variance of results across a system as the effect from individual noisy qubits becomes less appreciable, and layouts within a covering set increasingly overlap.  It might also be assumed that the device wide average over a covering set remains relatively constant when the number of qubits increases.  However, our 5-qubit DTC results are greatly impacted by edge-effects due to the short-length of the layouts.  In particular, qubits at the edge of a chain participate in two-qubit interactions with only a single neighboring qubit; one of the two terms in the denominator of Eq.~(\ref{eq:relation}) is zero.  In contrast, qubits in the bulk have two neighbors that they interact with and therefore, all else being equal, qubits at the interior of a layout have fewer visible cycles as they partake in more noisy interactions. We therefore expect that, at small numbers of qubits, the system-wide average visibility is boosted by edge-effects that are suppressed at larger layout lengths.
\begin{figure}[t]
\includegraphics[width=\linewidth]{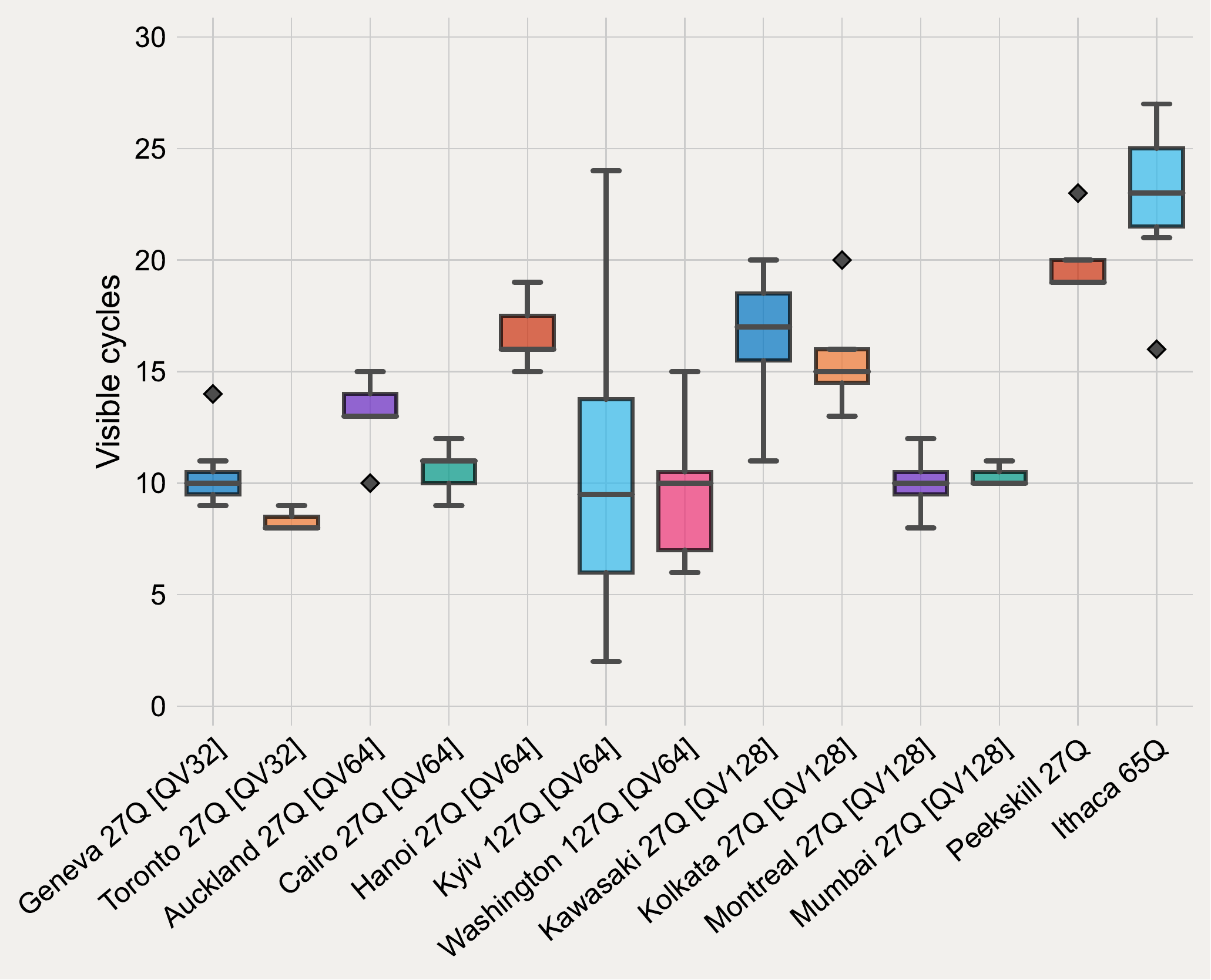}
\caption{Results of 20-qubit DTC covering set characterization.  Systems are ordered by increasing qubit count and reported QV values.}
\label{fig:20q_dtc}
\end{figure}

This intuition is confirmed in Fig.~(\ref{fig:20q_dtc}) where 20Q DTC characterization is performed over IBM Quantum systems with 27 or more qubits.  We see that there is an overall decrease in the average number of visible cycles over the covering sets as compared to our 5-qubit results, highlighting the edge-effects present in small width DTC circuits.  The decreased variance between layouts within a covering set is also inline with expectations, however at 20-qubits there are still large variations in performance across the IBM Quantum Kyiv system that are not seen on the Washington system of the same size.  This indicates that there are uncommonly large portions of the system exhibiting bipolar performance characteristics.  Once again, we see little correlation with reported QV values, with QV128 systems Montreal and Mumbai scoring on par with QV32 (Geneva) and QV64 (Cairo) machines.  While IBM Quantum Peekskill was the best performing system on 5-qubit DTC characterization, at 20-qubits the Ithaca system supports the largest average number of visible cycles.  Surprisingly, the average 5- and 20-qubit DTC scores are nearly identical for Ithaca, suggesting that edge-effects do not have a big impact on this system.  However, because the 20-qubit DTC data was not taken at the same time as the 5-qubit results, other issues such as device stability [Sec.~(\ref{sec:results5})] may also play an attributing role.

In Table~(\ref{tab:20Q+ results}) we focus on the 65Q IBM Quantum Ithaca and 127Q Washington systems going out to increasing numbers of qubits for DTC characterization.
\begin{table}[b]
    \begin{subtable}[h!]{0.5\textwidth}
    \centering
    \begin{tabular}{|c|c|c|c|c|}
    \hline
    & 20-qubit & 30-qubit & 40-qubit & 50-qubit \\
    \hline
    Mean cycle visibility & 22.7 & 25.0 & 25.6 & 25.5\\
    \hline
    Standard deviation & 3.5 & 3.0 & 1.6 & 1.1\\
    \hline
    Number of layouts & 6 & 4 & 4 & 4\\
    \hline 
    \end{tabular}
    \vspace{-0.05in}
    \caption{65Q Ithaca}
    \end{subtable}
    \begin{subtable}[h!]{0.5\textwidth}
    \centering 
    \vspace{0.1in}
    \begin{tabular}{|c|c|c|c|c|}
    \hline
    & 20-qubit & 40-qubit & 60-qubit & 80-qubit \\
    \hline
    Mean cycle visibility & 10.0 & 10.4 & 11.1 & 11.2\\
    \hline
    Standard deviation & 2.6 & 1.8 & 1.4 & 1.0\\
    \hline
    Number of layouts & 11 & 7 & 6 & 5\\
    \hline
    \end{tabular}
    \vspace{-0.05in}
    \caption{127Q Washington}
    \end{subtable}
    \caption{Mean cycle visibility, standard deviation, and number of layouts for the 65-qubit Ithaca and 127-qubit Washington systems as a function of the number of qubits used in DTC characterization.}
    \label{tab:20Q+ results}
\end{table}
The lack of appreciable changes in the number of visible cycles shown in Table.~(\ref{tab:20Q+ results}) as we increase the number of qubits indicates that DTC characterization at $\sim 20$ qubits is sufficient to suppress edge-effects in our covering set results.  Although increasing the length of the layouts used in DTC characterization reduces the variance in the data, it is clear even at 20-qubits that the Ithaca system performs much better than Washington, averaging $\sim 2.5$x the number of visible cycles; there is a pronounced reduction in the local noise seen by qubits in the Ithaca system as compared to the Washington device. This is generally inline with system calibration data taken from the machines, however differences in the relative performance between DTC characterization and predicted performance based on system calibration data occur across both machines, allowing only a qualitative comparison to be made. 

\section{Conclusion}\label{sec:conclusion}

We have demonstrated that DTC allows for meaningful, in-depth characterization of near-term quantum processor performance through systematic evaluation of the majority of the IBM Quantum processor fleet.  Our method is easy to evaluate, and can be executed across the full-topology of a quantum system, yielding insights into device quality that would otherwise go overlooked with single subgraph evaluation.  Owing to many-body noise-localization present in DTC, we are able to not only evaluate linear-chains of qubits but also pinpoint poor performing regions of a quantum system down to the single-qubit level.  Our results are corroborated by device calibration data, while providing a more faithful tally of system performance than is possible with one- and two-qubit benchmarking.  Using DTC we have tracked the stability of quantum processors, and have seen that there are sizable variations in absolute system performance, but that relative performance of qubit layouts does not vary greatly over time.  This indicates that metrics computed infrequently are not faithful indicators of performance, and we have shown that system quality as measured by DTC is in general not correlated with the reported QV values for IBM Quantum systems due to infrequent computation of the latter.

The noise-locality of DTC allows for the method described here to be extended in a variety of novel ways.  Rather than full-characterization, DTC can be used to efficiently probe a quantum system for ``faulty qubits"; qubits that can perform only a limited number of visible cycles.  Because the number of cycles for faulty qubits is small, the number of circuit executions is minimal, leading to a quick method by which regions of a system defined by the layouts in a covering set possessing large local noise can be interrogated.  Taken a step further, because each qubit acts as a local noise barometer, the number of visible cycles from DTC characterization is effectively mapping out the noise profile of a quantum device.  Thus with judicious selection of covering sets and readout error mitigation \cite{nation:2021}, it is possible to use the method here to create an effective error map of a quantum system.  The use of such error maps, either independently or in conjunction with with device calibration data, for qubit selection (e.g. using \texttt{mapomatic} \cite{nation:2022}) is currently an ongoing effort.  The wealth of information obtainable via DTC experiments, that goes beyond that found in other application-based test circuits, will likely make it a mainstay of quantum computing characterization methods for near-term quantum systems.

\begin{acknowledgments}
The authors thank Isaac Lauer, Doug McClure, David McKay, Hasan Nayfeh, and  Sami Rosenblatt for helpful discussions.  VZ was supported by a Quantum Undergraduate Research at IBM and Princeton (QURIP) 2022 internship.  All figures, save for Fig.~(\ref{fig:error_map}), are created with Matplotlib \cite{matplotlib}.
\end{acknowledgments}

\appendix

\begin{figure}[b]
\includegraphics[width=\linewidth]{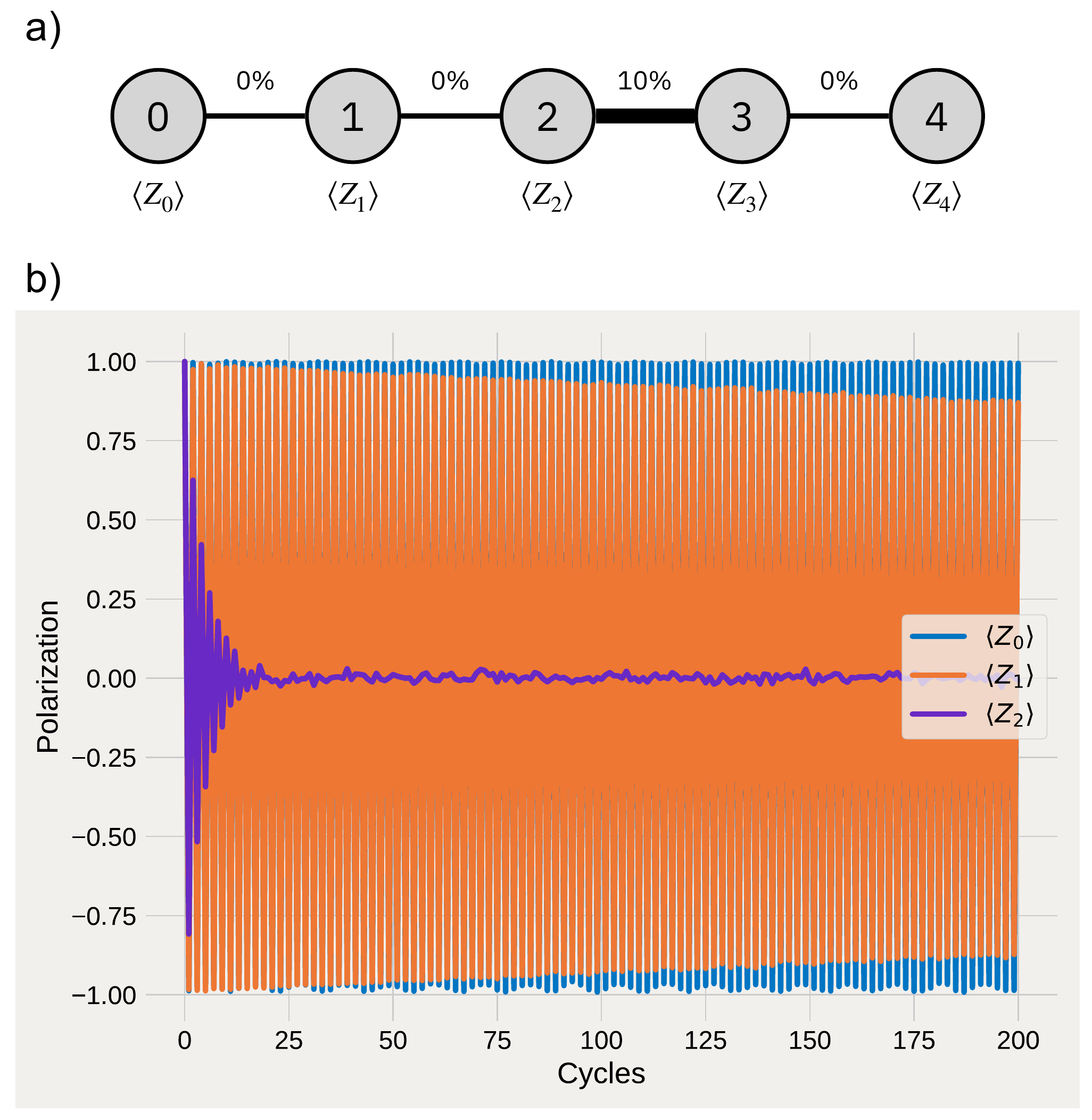}
\caption{(a) A linear nearest-neighbor chain of qubits with the CNOT error rate between qubits $2$ and $3$ set at $10\%$ assuming a depolarizing noise model.  (b) Simulation of the circuit from Fig.~(\ref{fig:circ}) using the model from (a) to compute the single-qubit polarization values for qubits $0\rightarrow 2$.  The results for qubits $3$ and $4$ closely match those of qubit $2$ and $1$, respectively. } 
\label{fig:locality}
\end{figure}

\section{Noise localization}\label{app:noise}
One of the advantages of utilizing DTC for characterization lies in the confinement of noise in many-body localized systems \cite{ippoliti:2021,serbyn:2013,bardarson:2012}.  This allows us to consider the number of visible cycles on each qubit as a relative gauge of the local noise strength.  Although this effect has already been examined in Ref.~\cite{ippoliti:2021}, we do so again here for completeness, and to confirm that this localization of noise is valid in the circuits used in this work.  To do so, we consider a 5-qubit linear nearest-neighbor topology, as shown in Fig.~(\ref{fig:locality}a), where the CNOT gate between qubits $Q2$ and $Q3$ is subject to a depolarizing noise model with $10\%$ error rate.  All other gate and qubit values are assumed to be ideal.  Figure~(\ref{fig:locality}b) shows simulations of the single-qubit polarization values, $\langle Z_{i}\rangle$, for qubits $0\rightarrow 2$ over a range of cycles for the circuit shown in Fig.~(\ref{fig:circ}).  From this figure it is clear that while the large error between $Q2$ and $Q3$ leads to a strong decay in the polarization for those qubits, other qubits not directly participating in the noisy two-qubit gate are only weakly affected.  In this example, although $Q2$ is heavily suppressed (only 5 visible cycles), we must go out relatively large numbers of cycles to see appreciable decay in the polarization for $Q1$.  $Q0$, being further away in the topology, shows even weaker signatures of the noise on its polarization values.  In Fig.~(\ref{fig:locality}b) both $Q1$ and $Q0$ have $200$ visible cycles, and could support many more had we considered a larger $N_{\rm max}$ in the simulations.

\bibliography{refs}

\begin{thebibliography}{29}%
\makeatletter
\providecommand \@ifxundefined [1]{%
 \@ifx{#1\undefined}
}%
\providecommand \@ifnum [1]{%
 \ifnum #1\expandafter \@firstoftwo
 \else \expandafter \@secondoftwo
 \fi
}%
\providecommand \@ifx [1]{%
 \ifx #1\expandafter \@firstoftwo
 \else \expandafter \@secondoftwo
 \fi
}%
\providecommand \natexlab [1]{#1}%
\providecommand \enquote  [1]{``#1''}%
\providecommand \bibnamefont  [1]{#1}%
\providecommand \bibfnamefont [1]{#1}%
\providecommand \citenamefont [1]{#1}%
\providecommand \href@noop [0]{\@secondoftwo}%
\providecommand \href [0]{\begingroup \@sanitize@url \@href}%
\providecommand \@href[1]{\@@startlink{#1}\@@href}%
\providecommand \@@href[1]{\endgroup#1\@@endlink}%
\providecommand \@sanitize@url [0]{\catcode `\\12\catcode `\$12\catcode
  `\&12\catcode `\#12\catcode `\^12\catcode `\_12\catcode `\%12\relax}%
\providecommand \@@startlink[1]{}%
\providecommand \@@endlink[0]{}%
\providecommand \url  [0]{\begingroup\@sanitize@url \@url }%
\providecommand \@url [1]{\endgroup\@href {#1}{\urlprefix }}%
\providecommand \urlprefix  [0]{URL }%
\providecommand \Eprint [0]{\href }%
\providecommand \doibase [0]{https://doi.org/}%
\providecommand \selectlanguage [0]{\@gobble}%
\providecommand \bibinfo  [0]{\@secondoftwo}%
\providecommand \bibfield  [0]{\@secondoftwo}%
\providecommand \translation [1]{[#1]}%
\providecommand \BibitemOpen [0]{}%
\providecommand \bibitemStop [0]{}%
\providecommand \bibitemNoStop [0]{.\EOS\space}%
\providecommand \EOS [0]{\spacefactor3000\relax}%
\providecommand \BibitemShut  [1]{\csname bibitem#1\endcsname}%
\let\auto@bib@innerbib\@empty
\bibitem [{\citenamefont {Magesan}\ \emph
  {et~al.}(2012{\natexlab{a}})\citenamefont {Magesan}, \citenamefont
  {Gambetta}, \citenamefont {Johnson}, \citenamefont {Ryan}, \citenamefont
  {Chow}, \citenamefont {Merkel}, \citenamefont {da~Silva}, \citenamefont
  {Keefe}, \citenamefont {Rothwell}, \citenamefont {Ohki}, \citenamefont
  {Ketchen},\ and\ \citenamefont {Steffen}}]{magesan:2012b}%
  \BibitemOpen
  \bibfield  {author} {\bibinfo {author} {\bibfnamefont {E.}~\bibnamefont
  {Magesan}}, \bibinfo {author} {\bibfnamefont {J.~M.}\ \bibnamefont
  {Gambetta}}, \bibinfo {author} {\bibfnamefont {B.~R.}\ \bibnamefont
  {Johnson}}, \bibinfo {author} {\bibfnamefont {C.~A.}\ \bibnamefont {Ryan}},
  \bibinfo {author} {\bibfnamefont {J.~M.}\ \bibnamefont {Chow}}, \bibinfo
  {author} {\bibfnamefont {S.~T.}\ \bibnamefont {Merkel}}, \bibinfo {author}
  {\bibfnamefont {M.~P.}\ \bibnamefont {da~Silva}}, \bibinfo {author}
  {\bibfnamefont {G.~A.}\ \bibnamefont {Keefe}}, \bibinfo {author}
  {\bibfnamefont {M.~B.}\ \bibnamefont {Rothwell}}, \bibinfo {author}
  {\bibfnamefont {T.~A.}\ \bibnamefont {Ohki}}, \bibinfo {author}
  {\bibfnamefont {M.~B.}\ \bibnamefont {Ketchen}},\ and\ \bibinfo {author}
  {\bibfnamefont {M.}~\bibnamefont {Steffen}},\ }\bibfield  {title} {\bibinfo
  {title} {{E}fficient {M}easurement of {Q}uantum {G}ate {E}rror by
  {I}nterleaved {R}andomized {B}enchmarking},\ }\href
  {https://doi.org/10.1103/PhysRevLett.109.080505} {\bibfield  {journal}
  {\bibinfo  {journal} {Phys. Rev. Lett.}\ }\textbf {\bibinfo {volume} {109}},\
  \bibinfo {pages} {080505} (\bibinfo {year} {2012}{\natexlab{a}})}\BibitemShut
  {NoStop}%
\bibitem [{\citenamefont {Magesan}\ \emph
  {et~al.}(2012{\natexlab{b}})\citenamefont {Magesan}, \citenamefont
  {Gambetta},\ and\ \citenamefont {Emerson}}]{magesan:2012}%
  \BibitemOpen
  \bibfield  {author} {\bibinfo {author} {\bibfnamefont {E.}~\bibnamefont
  {Magesan}}, \bibinfo {author} {\bibfnamefont {J.~M.}\ \bibnamefont
  {Gambetta}},\ and\ \bibinfo {author} {\bibfnamefont {J.}~\bibnamefont
  {Emerson}},\ }\bibfield  {title} {\bibinfo {title} {{C}haracterizing quantum
  gates via randomized benchmarking},\ }\href
  {https://doi.org/10.1103/PhysRevA.85.042311} {\bibfield  {journal} {\bibinfo
  {journal} {Phys. Rev. A}\ }\textbf {\bibinfo {volume} {85}},\ \bibinfo
  {pages} {042311} (\bibinfo {year} {2012}{\natexlab{b}})}\BibitemShut
  {NoStop}%
\bibitem [{\citenamefont {Magesan}\ \emph {et~al.}(2011)\citenamefont
  {Magesan}, \citenamefont {Gambetta},\ and\ \citenamefont
  {Emerson}}]{magesan:2011}%
  \BibitemOpen
  \bibfield  {author} {\bibinfo {author} {\bibfnamefont {E.}~\bibnamefont
  {Magesan}}, \bibinfo {author} {\bibfnamefont {J.~M.}\ \bibnamefont
  {Gambetta}},\ and\ \bibinfo {author} {\bibfnamefont {J.}~\bibnamefont
  {Emerson}},\ }\bibfield  {title} {\bibinfo {title} {{S}calable and {R}obust
  {R}andomized {B}enchmarking of {Q}uantum {P}rocesses},\ }\href
  {https://doi.org/10.1103/PhysRevLett.106.180504} {\bibfield  {journal}
  {\bibinfo  {journal} {Phys. Rev. Lett.}\ }\textbf {\bibinfo {volume} {106}},\
  \bibinfo {pages} {180504} (\bibinfo {year} {2011})}\BibitemShut {NoStop}%
\bibitem [{\citenamefont {Knill}\ \emph {et~al.}(2008)\citenamefont {Knill},
  \citenamefont {Leibfried}, \citenamefont {Reichle}, \citenamefont {Britton},
  \citenamefont {Blakestad}, \citenamefont {Jost}, \citenamefont {Langer},
  \citenamefont {Ozeri}, \citenamefont {Seidelin},\ and\ \citenamefont
  {Wineland}}]{knill:2008}%
  \BibitemOpen
  \bibfield  {author} {\bibinfo {author} {\bibfnamefont {E.}~\bibnamefont
  {Knill}}, \bibinfo {author} {\bibfnamefont {D.}~\bibnamefont {Leibfried}},
  \bibinfo {author} {\bibfnamefont {R.}~\bibnamefont {Reichle}}, \bibinfo
  {author} {\bibfnamefont {J.}~\bibnamefont {Britton}}, \bibinfo {author}
  {\bibfnamefont {R.~B.}\ \bibnamefont {Blakestad}}, \bibinfo {author}
  {\bibfnamefont {J.~D.}\ \bibnamefont {Jost}}, \bibinfo {author}
  {\bibfnamefont {C.}~\bibnamefont {Langer}}, \bibinfo {author} {\bibfnamefont
  {R.}~\bibnamefont {Ozeri}}, \bibinfo {author} {\bibfnamefont
  {S.}~\bibnamefont {Seidelin}},\ and\ \bibinfo {author} {\bibfnamefont
  {D.~J.}\ \bibnamefont {Wineland}},\ }\bibfield  {title} {\bibinfo {title}
  {{R}andomized benchmarking of quantum gates},\ }\href
  {https://doi.org/10.1103/PhysRevA.77.012307} {\bibfield  {journal} {\bibinfo
  {journal} {Phys. Rev. A}\ }\textbf {\bibinfo {volume} {77}},\ \bibinfo
  {pages} {012307} (\bibinfo {year} {2008})}\BibitemShut {NoStop}%
\bibitem [{\citenamefont {Proctor}\ \emph
  {et~al.}(2022{\natexlab{a}})\citenamefont {Proctor}, \citenamefont {Seritan},
  \citenamefont {Rudinger}, \citenamefont {Nielsen}, \citenamefont
  {Blume-Kohout},\ and\ \citenamefont {Young}}]{proctor:2022b}%
  \BibitemOpen
  \bibfield  {author} {\bibinfo {author} {\bibfnamefont {T.}~\bibnamefont
  {Proctor}}, \bibinfo {author} {\bibfnamefont {S.}~\bibnamefont {Seritan}},
  \bibinfo {author} {\bibfnamefont {K.}~\bibnamefont {Rudinger}}, \bibinfo
  {author} {\bibfnamefont {E.}~\bibnamefont {Nielsen}}, \bibinfo {author}
  {\bibfnamefont {R.}~\bibnamefont {Blume-Kohout}},\ and\ \bibinfo {author}
  {\bibfnamefont {K.}~\bibnamefont {Young}},\ }\bibfield  {title} {\bibinfo
  {title} {{S}calable {R}andomized {B}enchmarking of {Q}uantum {C}omputers
  {U}sing {M}irror {C}ircuits},\ }\href
  {https://doi.org/10.1103/PhysRevLett.129.150502} {\bibfield  {journal}
  {\bibinfo  {journal} {Phys. Rev. Lett.}\ }\textbf {\bibinfo {volume} {129}},\
  \bibinfo {pages} {150502} (\bibinfo {year} {2022}{\natexlab{a}})}\BibitemShut
  {NoStop}%
\bibitem [{\citenamefont {Cross}\ \emph {et~al.}(2019)\citenamefont {Cross},
  \citenamefont {Bishop}, \citenamefont {Sheldon}, \citenamefont {Nation},\
  and\ \citenamefont {Gambetta}}]{cross:2019}%
  \BibitemOpen
  \bibfield  {author} {\bibinfo {author} {\bibfnamefont {A.~W.}\ \bibnamefont
  {Cross}}, \bibinfo {author} {\bibfnamefont {L.~S.}\ \bibnamefont {Bishop}},
  \bibinfo {author} {\bibfnamefont {S.}~\bibnamefont {Sheldon}}, \bibinfo
  {author} {\bibfnamefont {P.~D.}\ \bibnamefont {Nation}},\ and\ \bibinfo
  {author} {\bibfnamefont {J.~M.}\ \bibnamefont {Gambetta}},\ }\bibfield
  {title} {\bibinfo {title} {{V}alidating quantum computers using randomized
  model circuits},\ }\href {https://doi.org/10.1103/PhysRevA.100.032328}
  {\bibfield  {journal} {\bibinfo  {journal} {Phys. Rev. A}\ }\textbf {\bibinfo
  {volume} {100}},\ \bibinfo {pages} {032328} (\bibinfo {year}
  {2019})}\BibitemShut {NoStop}%
\bibitem [{\citenamefont {Proctor}\ \emph
  {et~al.}(2022{\natexlab{b}})\citenamefont {Proctor}, \citenamefont
  {Rudinger}, \citenamefont {Young}, \citenamefont {Nielsen},\ and\
  \citenamefont {Blume-Kohout}}]{proctor:2022}%
  \BibitemOpen
  \bibfield  {author} {\bibinfo {author} {\bibfnamefont {T.}~\bibnamefont
  {Proctor}}, \bibinfo {author} {\bibfnamefont {K.}~\bibnamefont {Rudinger}},
  \bibinfo {author} {\bibfnamefont {K.}~\bibnamefont {Young}}, \bibinfo
  {author} {\bibfnamefont {E.}~\bibnamefont {Nielsen}},\ and\ \bibinfo {author}
  {\bibfnamefont {R.}~\bibnamefont {Blume-Kohout}},\ }\bibfield  {title}
  {\bibinfo {title} {{M}easuring the capabilities of quantum computers},\
  }\href {https://doi.org/10.1038/s41567-021-01409-7} {\bibfield  {journal}
  {\bibinfo  {journal} {Nat. Phys.}\ }\textbf {\bibinfo {volume} {18}},\
  \bibinfo {pages} {75} (\bibinfo {year} {2022}{\natexlab{b}})}\BibitemShut
  {NoStop}%
\bibitem [{\citenamefont {Tomesh}\ \emph {et~al.}(2022)\citenamefont {Tomesh},
  \citenamefont {Gokhale}, \citenamefont {Omole}, \citenamefont {Ravi},
  \citenamefont {Smith}, \citenamefont {Viszlai}, \citenamefont {Wu},
  \citenamefont {Hardavellas}, \citenamefont {Martonosi},\ and\ \citenamefont
  {Chong}}]{tomesh:2022}%
  \BibitemOpen
  \bibfield  {author} {\bibinfo {author} {\bibfnamefont {T.}~\bibnamefont
  {Tomesh}}, \bibinfo {author} {\bibfnamefont {P.}~\bibnamefont {Gokhale}},
  \bibinfo {author} {\bibfnamefont {V.}~\bibnamefont {Omole}}, \bibinfo
  {author} {\bibfnamefont {G.~S.}\ \bibnamefont {Ravi}}, \bibinfo {author}
  {\bibfnamefont {K.~N.}\ \bibnamefont {Smith}}, \bibinfo {author}
  {\bibfnamefont {J.}~\bibnamefont {Viszlai}}, \bibinfo {author} {\bibfnamefont
  {X.-C.}\ \bibnamefont {Wu}}, \bibinfo {author} {\bibfnamefont
  {N.}~\bibnamefont {Hardavellas}}, \bibinfo {author} {\bibfnamefont {M.~R.}\
  \bibnamefont {Martonosi}},\ and\ \bibinfo {author} {\bibfnamefont {F.~T.}\
  \bibnamefont {Chong}},\ }\bibfield  {title} {\bibinfo {title} {{S}upermar{Q}:
  {A} {S}calable {Q}uantum {B}enchmark {S}uite},\ }\bibfield  {journal}
  {\bibinfo  {journal} {arXiv:2202.11045}\ }\href
  {https://doi.org/10.48550/arXiv.2202.11045} {10.48550/arXiv.2202.11045}
  (\bibinfo {year} {2022})\BibitemShut {NoStop}%
\bibitem [{\citenamefont {Lubinski}\ \emph {et~al.}(2021)\citenamefont
  {Lubinski}, \citenamefont {Johri}, \citenamefont {Varosy}, \citenamefont
  {Coleman}, \citenamefont {Zhao}, \citenamefont {Necaise}, \citenamefont
  {Baldwin}, \citenamefont {Mayer},\ and\ \citenamefont
  {Proctor}}]{lubinsky:2021}%
  \BibitemOpen
  \bibfield  {author} {\bibinfo {author} {\bibfnamefont {T.}~\bibnamefont
  {Lubinski}}, \bibinfo {author} {\bibfnamefont {S.}~\bibnamefont {Johri}},
  \bibinfo {author} {\bibfnamefont {P.}~\bibnamefont {Varosy}}, \bibinfo
  {author} {\bibfnamefont {J.}~\bibnamefont {Coleman}}, \bibinfo {author}
  {\bibfnamefont {L.}~\bibnamefont {Zhao}}, \bibinfo {author} {\bibfnamefont
  {J.}~\bibnamefont {Necaise}}, \bibinfo {author} {\bibfnamefont {C.~H.}\
  \bibnamefont {Baldwin}}, \bibinfo {author} {\bibfnamefont {K.}~\bibnamefont
  {Mayer}},\ and\ \bibinfo {author} {\bibfnamefont {T.}~\bibnamefont
  {Proctor}},\ }\bibfield  {title} {\bibinfo {title} {{A}pplication-{O}riented
  {P}erformance {B}enchmarks for {Q}uantum {C}omputing},\ }\bibfield  {journal}
  {\bibinfo  {journal} {arXiv:2110.03137}\ }\href
  {https://doi.org/10.48550/arXiv.2110.03137} {10.48550/arXiv.2110.03137}
  (\bibinfo {year} {2021})\BibitemShut {NoStop}%
\bibitem [{\citenamefont {Blume-Kohout}\ and\ \citenamefont
  {Young}(2020)}]{kohout:2020}%
  \BibitemOpen
  \bibfield  {author} {\bibinfo {author} {\bibfnamefont {R.}~\bibnamefont
  {Blume-Kohout}}\ and\ \bibinfo {author} {\bibfnamefont {K.~C.}\ \bibnamefont
  {Young}},\ }\bibfield  {title} {\bibinfo {title} {{A} volumetric framework
  for quantum computer benchmarks},\ }\href
  {https://doi.org/10.22331/q-2020-11-15-362} {\bibfield  {journal} {\bibinfo
  {journal} {Quantum}\ }\textbf {\bibinfo {volume} {4}},\ \bibinfo {pages}
  {362} (\bibinfo {year} {2020})}\BibitemShut {NoStop}%
\bibitem [{\citenamefont {Li}\ \emph {et~al.}(2020)\citenamefont {Li},
  \citenamefont {Stein}, \citenamefont {Krishnamoorthy},\ and\ \citenamefont
  {Ang}}]{li:2020}%
  \BibitemOpen
  \bibfield  {author} {\bibinfo {author} {\bibfnamefont {A.}~\bibnamefont
  {Li}}, \bibinfo {author} {\bibfnamefont {S.}~\bibnamefont {Stein}}, \bibinfo
  {author} {\bibfnamefont {S.}~\bibnamefont {Krishnamoorthy}},\ and\ \bibinfo
  {author} {\bibfnamefont {J.}~\bibnamefont {Ang}},\ }\bibfield  {title}
  {\bibinfo {title} {{QASMB}ench: {A} {L}ow-level {QASM} {B}enchmark {S}uite
  for {NISQ} {E}valuation and {S}imulation},\ }\bibfield  {journal} {\bibinfo
  {journal} {arXiv:2005.13018}\ }\href
  {https://doi.org/10.48550/arXiv.2005.13018} {10.48550/arXiv.2005.13018}
  (\bibinfo {year} {2020})\BibitemShut {NoStop}%
\bibitem [{\citenamefont {Frey}\ and\ \citenamefont
  {Rachel}(2022)}]{frey:2022}%
  \BibitemOpen
  \bibfield  {author} {\bibinfo {author} {\bibfnamefont {P.}~\bibnamefont
  {Frey}}\ and\ \bibinfo {author} {\bibfnamefont {S.}~\bibnamefont {Rachel}},\
  }\bibfield  {title} {\bibinfo {title} {Realization of a discrete time crystal
  on 57 qubits of a quantum computer},\ }\href
  {https://doi.org/10.1126/sciadv.abm7652} {\bibfield  {journal} {\bibinfo
  {journal} {Science Advances}\ }\textbf {\bibinfo {volume} {8}},\ \bibinfo
  {pages} {eabm7652} (\bibinfo {year} {2022})}\BibitemShut {NoStop}%
\bibitem [{\citenamefont {Mi}\ \emph {et~al.}(2022)\citenamefont {Mi},
  \citenamefont {Ippoliti}, \citenamefont {Quintana}, \citenamefont {Greene},
  \citenamefont {Chen}, \citenamefont {Gross}, \citenamefont {Arute},
  \citenamefont {Arya}, \citenamefont {Atalaya}, \citenamefont {Babbush},
  \citenamefont {Bardin}, \citenamefont {Basso}, \citenamefont {Bengtsson},
  \citenamefont {Bilmes}, \citenamefont {Bourassa}, \citenamefont {Brill},
  \citenamefont {Broughton}, \citenamefont {Buckley}, \citenamefont {Buell},
  \citenamefont {Burkett}, \citenamefont {Bushnell}, \citenamefont {Chiaro},
  \citenamefont {Collins}, \citenamefont {Courtney}, \citenamefont {Debroy},
  \citenamefont {Demura}, \citenamefont {Derk}, \citenamefont {Dunsworth},
  \citenamefont {Eppens}, \citenamefont {Erickson}, \citenamefont {Farhi},
  \citenamefont {Fowler}, \citenamefont {Foxen}, \citenamefont {Gidney},
  \citenamefont {Giustina}, \citenamefont {Harrigan}, \citenamefont
  {Harrington}, \citenamefont {Hilton}, \citenamefont {Ho}, \citenamefont
  {Hong}, \citenamefont {Huang}, \citenamefont {Huff}, \citenamefont {Huggins},
  \citenamefont {Ioffe}, \citenamefont {Isakov}, \citenamefont {Iveland},
  \citenamefont {Jeffrey}, \citenamefont {Jiang}, \citenamefont {Jones},
  \citenamefont {Kafri}, \citenamefont {Khattar}, \citenamefont {Kim},
  \citenamefont {Kitaev}, \citenamefont {Klimov}, \citenamefont {Korotkov},
  \citenamefont {Kostritsa}, \citenamefont {Landhuis}, \citenamefont {Laptev},
  \citenamefont {Lee}, \citenamefont {Lee}, \citenamefont {Locharla},
  \citenamefont {Lucero}, \citenamefont {Martin}, \citenamefont {McClean},
  \citenamefont {McCourt}, \citenamefont {McEwen}, \citenamefont {Miao},
  \citenamefont {Mohseni}, \citenamefont {Montazeri}, \citenamefont
  {Mruczkiewicz}, \citenamefont {Naaman}, \citenamefont {Neeley}, \citenamefont
  {Neill}, \citenamefont {Newman}, \citenamefont {Niu}, \citenamefont
  {O'Brien}, \citenamefont {Opremcak}, \citenamefont {Ostby}, \citenamefont
  {Pato}, \citenamefont {Petukhov}, \citenamefont {Rubin}, \citenamefont
  {Sank}, \citenamefont {Satzinger}, \citenamefont {Shvarts}, \citenamefont
  {Su}, \citenamefont {Strain}, \citenamefont {Szalay}, \citenamefont
  {Trevithick}, \citenamefont {Villalonga}, \citenamefont {White},
  \citenamefont {Yao}, \citenamefont {Yeh}, \citenamefont {Yoo}, \citenamefont
  {Zalcman}, \citenamefont {Neven}, \citenamefont {Boixo}, \citenamefont
  {Smelyanskiy}, \citenamefont {Megrant}, \citenamefont {Kelly}, \citenamefont
  {Chen}, \citenamefont {Sondhi}, \citenamefont {Moessner}, \citenamefont
  {Kechedzhi}, \citenamefont {Khemani},\ and\ \citenamefont
  {Roushan}}]{mi:2022}%
  \BibitemOpen
  \bibfield  {author} {\bibinfo {author} {\bibfnamefont {X.}~\bibnamefont
  {Mi}}, \bibinfo {author} {\bibfnamefont {M.}~\bibnamefont {Ippoliti}},
  \bibinfo {author} {\bibfnamefont {C.}~\bibnamefont {Quintana}}, \bibinfo
  {author} {\bibfnamefont {A.}~\bibnamefont {Greene}}, \bibinfo {author}
  {\bibfnamefont {Z.}~\bibnamefont {Chen}}, \bibinfo {author} {\bibfnamefont
  {J.}~\bibnamefont {Gross}}, \bibinfo {author} {\bibfnamefont
  {F.}~\bibnamefont {Arute}}, \bibinfo {author} {\bibfnamefont
  {K.}~\bibnamefont {Arya}}, \bibinfo {author} {\bibfnamefont {J.}~\bibnamefont
  {Atalaya}}, \bibinfo {author} {\bibfnamefont {R.}~\bibnamefont {Babbush}},
  \bibinfo {author} {\bibfnamefont {J.~C.}\ \bibnamefont {Bardin}}, \bibinfo
  {author} {\bibfnamefont {J.}~\bibnamefont {Basso}}, \bibinfo {author}
  {\bibfnamefont {A.}~\bibnamefont {Bengtsson}}, \bibinfo {author}
  {\bibfnamefont {A.}~\bibnamefont {Bilmes}}, \bibinfo {author} {\bibfnamefont
  {A.}~\bibnamefont {Bourassa}}, \bibinfo {author} {\bibfnamefont
  {L.}~\bibnamefont {Brill}}, \bibinfo {author} {\bibfnamefont
  {M.}~\bibnamefont {Broughton}}, \bibinfo {author} {\bibfnamefont {B.~B.}\
  \bibnamefont {Buckley}}, \bibinfo {author} {\bibfnamefont {D.~A.}\
  \bibnamefont {Buell}}, \bibinfo {author} {\bibfnamefont {B.}~\bibnamefont
  {Burkett}}, \bibinfo {author} {\bibfnamefont {N.}~\bibnamefont {Bushnell}},
  \bibinfo {author} {\bibfnamefont {B.}~\bibnamefont {Chiaro}}, \bibinfo
  {author} {\bibfnamefont {R.}~\bibnamefont {Collins}}, \bibinfo {author}
  {\bibfnamefont {W.}~\bibnamefont {Courtney}}, \bibinfo {author}
  {\bibfnamefont {D.}~\bibnamefont {Debroy}}, \bibinfo {author} {\bibfnamefont
  {S.}~\bibnamefont {Demura}}, \bibinfo {author} {\bibfnamefont {A.~R.}\
  \bibnamefont {Derk}}, \bibinfo {author} {\bibfnamefont {A.}~\bibnamefont
  {Dunsworth}}, \bibinfo {author} {\bibfnamefont {D.}~\bibnamefont {Eppens}},
  \bibinfo {author} {\bibfnamefont {C.}~\bibnamefont {Erickson}}, \bibinfo
  {author} {\bibfnamefont {E.}~\bibnamefont {Farhi}}, \bibinfo {author}
  {\bibfnamefont {A.~G.}\ \bibnamefont {Fowler}}, \bibinfo {author}
  {\bibfnamefont {B.}~\bibnamefont {Foxen}}, \bibinfo {author} {\bibfnamefont
  {C.}~\bibnamefont {Gidney}}, \bibinfo {author} {\bibfnamefont
  {M.}~\bibnamefont {Giustina}}, \bibinfo {author} {\bibfnamefont {M.~P.}\
  \bibnamefont {Harrigan}}, \bibinfo {author} {\bibfnamefont {S.~D.}\
  \bibnamefont {Harrington}}, \bibinfo {author} {\bibfnamefont
  {J.}~\bibnamefont {Hilton}}, \bibinfo {author} {\bibfnamefont
  {A.}~\bibnamefont {Ho}}, \bibinfo {author} {\bibfnamefont {S.}~\bibnamefont
  {Hong}}, \bibinfo {author} {\bibfnamefont {T.}~\bibnamefont {Huang}},
  \bibinfo {author} {\bibfnamefont {A.}~\bibnamefont {Huff}}, \bibinfo {author}
  {\bibfnamefont {W.~J.}\ \bibnamefont {Huggins}}, \bibinfo {author}
  {\bibfnamefont {L.~B.}\ \bibnamefont {Ioffe}}, \bibinfo {author}
  {\bibfnamefont {S.~V.}\ \bibnamefont {Isakov}}, \bibinfo {author}
  {\bibfnamefont {J.}~\bibnamefont {Iveland}}, \bibinfo {author} {\bibfnamefont
  {E.}~\bibnamefont {Jeffrey}}, \bibinfo {author} {\bibfnamefont
  {Z.}~\bibnamefont {Jiang}}, \bibinfo {author} {\bibfnamefont
  {C.}~\bibnamefont {Jones}}, \bibinfo {author} {\bibfnamefont
  {D.}~\bibnamefont {Kafri}}, \bibinfo {author} {\bibfnamefont
  {T.}~\bibnamefont {Khattar}}, \bibinfo {author} {\bibfnamefont
  {S.}~\bibnamefont {Kim}}, \bibinfo {author} {\bibfnamefont {A.}~\bibnamefont
  {Kitaev}}, \bibinfo {author} {\bibfnamefont {P.~V.}\ \bibnamefont {Klimov}},
  \bibinfo {author} {\bibfnamefont {A.~N.}\ \bibnamefont {Korotkov}}, \bibinfo
  {author} {\bibfnamefont {F.}~\bibnamefont {Kostritsa}}, \bibinfo {author}
  {\bibfnamefont {D.}~\bibnamefont {Landhuis}}, \bibinfo {author}
  {\bibfnamefont {P.}~\bibnamefont {Laptev}}, \bibinfo {author} {\bibfnamefont
  {J.}~\bibnamefont {Lee}}, \bibinfo {author} {\bibfnamefont {K.}~\bibnamefont
  {Lee}}, \bibinfo {author} {\bibfnamefont {A.}~\bibnamefont {Locharla}},
  \bibinfo {author} {\bibfnamefont {E.}~\bibnamefont {Lucero}}, \bibinfo
  {author} {\bibfnamefont {O.}~\bibnamefont {Martin}}, \bibinfo {author}
  {\bibfnamefont {J.~R.}\ \bibnamefont {McClean}}, \bibinfo {author}
  {\bibfnamefont {T.}~\bibnamefont {McCourt}}, \bibinfo {author} {\bibfnamefont
  {M.}~\bibnamefont {McEwen}}, \bibinfo {author} {\bibfnamefont {K.~C.}\
  \bibnamefont {Miao}}, \bibinfo {author} {\bibfnamefont {M.}~\bibnamefont
  {Mohseni}}, \bibinfo {author} {\bibfnamefont {S.}~\bibnamefont {Montazeri}},
  \bibinfo {author} {\bibfnamefont {W.}~\bibnamefont {Mruczkiewicz}}, \bibinfo
  {author} {\bibfnamefont {O.}~\bibnamefont {Naaman}}, \bibinfo {author}
  {\bibfnamefont {M.}~\bibnamefont {Neeley}}, \bibinfo {author} {\bibfnamefont
  {C.}~\bibnamefont {Neill}}, \bibinfo {author} {\bibfnamefont
  {M.}~\bibnamefont {Newman}}, \bibinfo {author} {\bibfnamefont {M.~Y.}\
  \bibnamefont {Niu}}, \bibinfo {author} {\bibfnamefont {T.~E.}\ \bibnamefont
  {O'Brien}}, \bibinfo {author} {\bibfnamefont {A.}~\bibnamefont {Opremcak}},
  \bibinfo {author} {\bibfnamefont {E.}~\bibnamefont {Ostby}}, \bibinfo
  {author} {\bibfnamefont {B.}~\bibnamefont {Pato}}, \bibinfo {author}
  {\bibfnamefont {A.}~\bibnamefont {Petukhov}}, \bibinfo {author}
  {\bibfnamefont {N.~C.}\ \bibnamefont {Rubin}}, \bibinfo {author}
  {\bibfnamefont {D.}~\bibnamefont {Sank}}, \bibinfo {author} {\bibfnamefont
  {K.~J.}\ \bibnamefont {Satzinger}}, \bibinfo {author} {\bibfnamefont
  {V.}~\bibnamefont {Shvarts}}, \bibinfo {author} {\bibfnamefont
  {Y.}~\bibnamefont {Su}}, \bibinfo {author} {\bibfnamefont {D.}~\bibnamefont
  {Strain}}, \bibinfo {author} {\bibfnamefont {M.}~\bibnamefont {Szalay}},
  \bibinfo {author} {\bibfnamefont {M.~D.}\ \bibnamefont {Trevithick}},
  \bibinfo {author} {\bibfnamefont {B.}~\bibnamefont {Villalonga}}, \bibinfo
  {author} {\bibfnamefont {T.}~\bibnamefont {White}}, \bibinfo {author}
  {\bibfnamefont {Z.~J.}\ \bibnamefont {Yao}}, \bibinfo {author} {\bibfnamefont
  {P.}~\bibnamefont {Yeh}}, \bibinfo {author} {\bibfnamefont {J.}~\bibnamefont
  {Yoo}}, \bibinfo {author} {\bibfnamefont {A.}~\bibnamefont {Zalcman}},
  \bibinfo {author} {\bibfnamefont {H.}~\bibnamefont {Neven}}, \bibinfo
  {author} {\bibfnamefont {S.}~\bibnamefont {Boixo}}, \bibinfo {author}
  {\bibfnamefont {V.}~\bibnamefont {Smelyanskiy}}, \bibinfo {author}
  {\bibfnamefont {A.}~\bibnamefont {Megrant}}, \bibinfo {author} {\bibfnamefont
  {J.}~\bibnamefont {Kelly}}, \bibinfo {author} {\bibfnamefont
  {Y.}~\bibnamefont {Chen}}, \bibinfo {author} {\bibfnamefont {S.~L.}\
  \bibnamefont {Sondhi}}, \bibinfo {author} {\bibfnamefont {R.}~\bibnamefont
  {Moessner}}, \bibinfo {author} {\bibfnamefont {K.}~\bibnamefont {Kechedzhi}},
  \bibinfo {author} {\bibfnamefont {V.}~\bibnamefont {Khemani}},\ and\ \bibinfo
  {author} {\bibfnamefont {P.}~\bibnamefont {Roushan}},\ }\bibfield  {title}
  {\bibinfo {title} {Time-crystalline eigenstate order on a quantum
  processor},\ }\href {https://doi.org/10.1038/s41586-021-04257-w} {\bibfield
  {journal} {\bibinfo  {journal} {Nature}\ }\textbf {\bibinfo {volume} {601}},\
  \bibinfo {pages} {531} (\bibinfo {year} {2022})}\BibitemShut {NoStop}%
\bibitem [{\citenamefont {Wallman}\ and\ \citenamefont
  {Emerson}(2016)}]{wallman:2016}%
  \BibitemOpen
  \bibfield  {author} {\bibinfo {author} {\bibfnamefont {J.~J.}\ \bibnamefont
  {Wallman}}\ and\ \bibinfo {author} {\bibfnamefont {J.}~\bibnamefont
  {Emerson}},\ }\bibfield  {title} {\bibinfo {title} {{N}oise tailoring for
  scalable quantum computation via randomized compiling},\ }\href
  {https://doi.org/10.1103/PhysRevA.94.052325} {\bibfield  {journal} {\bibinfo
  {journal} {Phys. Rev. A}\ }\textbf {\bibinfo {volume} {94}},\ \bibinfo
  {pages} {052325} (\bibinfo {year} {2016})}\BibitemShut {NoStop}%
\bibitem [{\citenamefont {Knill}(2004)}]{knill:2004}%
  \BibitemOpen
  \bibfield  {author} {\bibinfo {author} {\bibfnamefont {E.}~\bibnamefont
  {Knill}},\ }\bibfield  {title} {\bibinfo {title} {{F}ault-{T}olerant
  {P}ostselected {Q}uantum {C}omputation: {T}hreshold {A}nalysis},\ }\bibfield
  {journal} {\bibinfo  {journal} {arXiv:0404104}\ }\href
  {https://doi.org/10.48550/arXiv.quant-ph/0404104}
  {10.48550/arXiv.quant-ph/0404104} (\bibinfo {year} {2004})\BibitemShut
  {NoStop}%
\bibitem [{\citenamefont {Nation}\ and\ \citenamefont
  {Treinish}(2022)}]{nation:2022}%
  \BibitemOpen
  \bibfield  {author} {\bibinfo {author} {\bibfnamefont {P.~D.}\ \bibnamefont
  {Nation}}\ and\ \bibinfo {author} {\bibfnamefont {M.}~\bibnamefont
  {Treinish}},\ }\bibfield  {title} {\bibinfo {title} {{S}uppressing quantum
  circuit errors due to system variability},\ }\bibfield  {journal} {\bibinfo
  {journal} {arXiv:2209.15512}\ }\href
  {https://doi.org/10.48550/arXiv.2209.15512} {10.48550/arXiv.2209.15512}
  (\bibinfo {year} {2022})\BibitemShut {NoStop}%
\bibitem [{\citenamefont {Ippoliti}\ \emph {et~al.}(2021)\citenamefont
  {Ippoliti}, \citenamefont {Kechedzhi}, \citenamefont {Moessner},
  \citenamefont {Sondhi},\ and\ \citenamefont {Khemani}}]{ippoliti:2021}%
  \BibitemOpen
  \bibfield  {author} {\bibinfo {author} {\bibfnamefont {M.}~\bibnamefont
  {Ippoliti}}, \bibinfo {author} {\bibfnamefont {K.}~\bibnamefont {Kechedzhi}},
  \bibinfo {author} {\bibfnamefont {R.}~\bibnamefont {Moessner}}, \bibinfo
  {author} {\bibfnamefont {S.~L.}\ \bibnamefont {Sondhi}},\ and\ \bibinfo
  {author} {\bibfnamefont {V.}~\bibnamefont {Khemani}},\ }\bibfield  {title}
  {\bibinfo {title} {{M}any-{B}ody {P}hysics in the {NISQ} {E}ra: {Q}uantum
  {P}rogramming a {D}iscrete {T}ime {C}rystal},\ }\href
  {https://doi.org/10.1103/PRXQuantum.2.030346} {\bibfield  {journal} {\bibinfo
   {journal} {PRX Quantum}\ }\textbf {\bibinfo {volume} {2}},\ \bibinfo {pages}
  {030346} (\bibinfo {year} {2021})}\BibitemShut {NoStop}%
\bibitem [{\citenamefont {Serbyn}\ \emph {et~al.}(2013)\citenamefont {Serbyn},
  \citenamefont {Papi\'{c}},\ and\ \citenamefont {Abanin}}]{serbyn:2013}%
  \BibitemOpen
  \bibfield  {author} {\bibinfo {author} {\bibfnamefont {M.}~\bibnamefont
  {Serbyn}}, \bibinfo {author} {\bibfnamefont {A.}~\bibnamefont {Papi\'{c}}},\
  and\ \bibinfo {author} {\bibfnamefont {D.~A.}\ \bibnamefont {Abanin}},\
  }\bibfield  {title} {\bibinfo {title} {{U}niversal {S}low {G}rowth of
  {E}ntanglement in {I}nteracting {S}trongly {D}isordered {S}ystems},\ }\href
  {https://doi.org/10.1103/PhysRevLett.110.260601} {\bibfield  {journal}
  {\bibinfo  {journal} {Phys. Rev. Lett.}\ }\textbf {\bibinfo {volume} {110}},\
  \bibinfo {pages} {260601} (\bibinfo {year} {2013})}\BibitemShut {NoStop}%
\bibitem [{\citenamefont {Bardarson}\ \emph {et~al.}(2012)\citenamefont
  {Bardarson}, \citenamefont {Pollmann},\ and\ \citenamefont
  {Moore}}]{bardarson:2012}%
  \BibitemOpen
  \bibfield  {author} {\bibinfo {author} {\bibfnamefont {J.~H.}\ \bibnamefont
  {Bardarson}}, \bibinfo {author} {\bibfnamefont {F.}~\bibnamefont
  {Pollmann}},\ and\ \bibinfo {author} {\bibfnamefont {J.~E.}\ \bibnamefont
  {Moore}},\ }\bibfield  {title} {\bibinfo {title} {{U}nbounded {G}rowth of
  {E}ntanglement in {M}odels of {M}any-{B}ody {L}ocalization},\ }\href
  {https://doi.org/10.1103/PhysRevLett.109.017202} {\bibfield  {journal}
  {\bibinfo  {journal} {Phys. Rev. Lett.}\ }\textbf {\bibinfo {volume} {109}},\
  \bibinfo {pages} {017202} (\bibinfo {year} {2012})}\BibitemShut {NoStop}%
\bibitem [{\citenamefont {Khemani}\ \emph {et~al.}(2016)\citenamefont
  {Khemani}, \citenamefont {Lazarides}, \citenamefont {Moessner},\ and\
  \citenamefont {Sondhi}}]{khemani:2016}%
  \BibitemOpen
  \bibfield  {author} {\bibinfo {author} {\bibfnamefont {V.}~\bibnamefont
  {Khemani}}, \bibinfo {author} {\bibfnamefont {A.}~\bibnamefont {Lazarides}},
  \bibinfo {author} {\bibfnamefont {R.}~\bibnamefont {Moessner}},\ and\
  \bibinfo {author} {\bibfnamefont {S.~L.}\ \bibnamefont {Sondhi}},\ }\bibfield
   {title} {\bibinfo {title} {{P}hase {S}tructure of {D}riven {Q}uantum
  {S}ystems},\ }\href {https://doi.org/10.1103/PhysRevLett.116.250401}
  {\bibfield  {journal} {\bibinfo  {journal} {Phys. Rev. Lett.}\ }\textbf
  {\bibinfo {volume} {116}},\ \bibinfo {pages} {250401} (\bibinfo {year}
  {2016})}\BibitemShut {NoStop}%
\bibitem [{\citenamefont {Treinish}\ \emph {et~al.}(2021)\citenamefont
  {Treinish}, \citenamefont {Carvalho}, \citenamefont {Tsilimigkounakis},\ and\
  \citenamefont {S\'{a}}}]{treinish:2021}%
  \BibitemOpen
  \bibfield  {author} {\bibinfo {author} {\bibfnamefont {M.}~\bibnamefont
  {Treinish}}, \bibinfo {author} {\bibfnamefont {I.}~\bibnamefont {Carvalho}},
  \bibinfo {author} {\bibfnamefont {G.}~\bibnamefont {Tsilimigkounakis}},\ and\
  \bibinfo {author} {\bibfnamefont {N.}~\bibnamefont {S\'{a}}},\ }\bibfield
  {title} {\bibinfo {title} {rustworkx: {A} {H}igh-{P}erformance {G}raph
  {L}ibrary for {P}ython},\ }\bibfield  {journal} {\bibinfo  {journal}
  {arXiv:2110.15221}\ }\href {https://doi.org/10.48550/arXiv.2110.15221}
  {10.48550/arXiv.2110.15221} (\bibinfo {year} {2021})\BibitemShut {NoStop}%
\bibitem [{\citenamefont {J{\"u}ttner}\ and\ \citenamefont
  {Madarasi}(2018)}]{vf2++}%
  \BibitemOpen
  \bibfield  {author} {\bibinfo {author} {\bibfnamefont {A.}~\bibnamefont
  {J{\"u}ttner}}\ and\ \bibinfo {author} {\bibfnamefont {P.}~\bibnamefont
  {Madarasi}},\ }\bibfield  {title} {\bibinfo {title} {Vf2++---an improved
  subgraph isomorphism algorithm},\ }\href
  {https://doi.org/https://doi.org/10.1016/j.dam.2018.02.018} {\bibfield
  {journal} {\bibinfo  {journal} {Discrete Applied Mathematics}\ }\textbf
  {\bibinfo {volume} {242}},\ \bibinfo {pages} {69} (\bibinfo {year} {2018})},\
  \bibinfo {note} {computational Advances in Combinatorial
  Optimization}\BibitemShut {NoStop}%
\bibitem [{rus(2022)}]{rustworkx}%
  \BibitemOpen
  \href {https://github.com/Qiskit/rustworkx} {\bibinfo {title}
  {https://github.com/qiskit/rustworkx}} (\bibinfo {year} {2022})\BibitemShut
  {NoStop}%
\bibitem [{cry(2022)}]{crystalmark}%
  \BibitemOpen
  \href {https://github.com/IBM-Quantum-Technical-Enablement/crystalmark}
  {\bibinfo {title}
  {https://github.com/ibm-quantum-technical-enablement/crystalmark}} (\bibinfo
  {year} {2022})\BibitemShut {NoStop}%
\bibitem [{qis(2022)}]{qiskit}%
  \BibitemOpen
  \href {https://qiskit.org} {\bibinfo {title} {Qiskit, https://qiskit.org}}
  (\bibinfo {year} {2022})\BibitemShut {NoStop}%
\bibitem [{map(2022)}]{mapo}%
  \BibitemOpen
  \href {https://github.com/Qiskit-Partners/mapomatic} {\bibinfo {title}
  {https://github.com/qiskit-partners/mapomatic}} (\bibinfo {year}
  {2022})\BibitemShut {NoStop}%
\bibitem [{\citenamefont {Jurcevic}\ \emph {et~al.}(2021)\citenamefont
  {Jurcevic}, \citenamefont {Javadi-Abhari}, \citenamefont {Bishop},
  \citenamefont {Lauer}, \citenamefont {Bogorin}, \citenamefont {Brink},
  \citenamefont {Capelluto}, \citenamefont {G{\"u}nl{\"u}k}, \citenamefont
  {Itoko}, \citenamefont {Kanazawa}, \citenamefont {Kandala}, \citenamefont
  {Keefe}, \citenamefont {Krsulich}, \citenamefont {Landers}, \citenamefont
  {Lewandowski}, \citenamefont {McClure}, \citenamefont {Nannicini},
  \citenamefont {Narasgond}, \citenamefont {Nayfeh}, \citenamefont {Pritchett},
  \citenamefont {Rothwell}, \citenamefont {Srinivasan}, \citenamefont
  {Sundaresan}, \citenamefont {Wang}, \citenamefont {Wei}, \citenamefont
  {Wood}, \citenamefont {Yau}, \citenamefont {Zhang}, \citenamefont {Dial},
  \citenamefont {Chow},\ and\ \citenamefont {Gambetta}}]{jurcevic:2021}%
  \BibitemOpen
  \bibfield  {author} {\bibinfo {author} {\bibfnamefont {P.}~\bibnamefont
  {Jurcevic}}, \bibinfo {author} {\bibfnamefont {A.}~\bibnamefont
  {Javadi-Abhari}}, \bibinfo {author} {\bibfnamefont {L.~S.}\ \bibnamefont
  {Bishop}}, \bibinfo {author} {\bibfnamefont {I.}~\bibnamefont {Lauer}},
  \bibinfo {author} {\bibfnamefont {D.~F.}\ \bibnamefont {Bogorin}}, \bibinfo
  {author} {\bibfnamefont {M.}~\bibnamefont {Brink}}, \bibinfo {author}
  {\bibfnamefont {L.}~\bibnamefont {Capelluto}}, \bibinfo {author}
  {\bibfnamefont {O.}~\bibnamefont {G{\"u}nl{\"u}k}}, \bibinfo {author}
  {\bibfnamefont {T.}~\bibnamefont {Itoko}}, \bibinfo {author} {\bibfnamefont
  {N.}~\bibnamefont {Kanazawa}}, \bibinfo {author} {\bibfnamefont
  {A.}~\bibnamefont {Kandala}}, \bibinfo {author} {\bibfnamefont {G.~A.}\
  \bibnamefont {Keefe}}, \bibinfo {author} {\bibfnamefont {K.}~\bibnamefont
  {Krsulich}}, \bibinfo {author} {\bibfnamefont {W.}~\bibnamefont {Landers}},
  \bibinfo {author} {\bibfnamefont {E.~P.}\ \bibnamefont {Lewandowski}},
  \bibinfo {author} {\bibfnamefont {D.~T.}\ \bibnamefont {McClure}}, \bibinfo
  {author} {\bibfnamefont {G.}~\bibnamefont {Nannicini}}, \bibinfo {author}
  {\bibfnamefont {A.}~\bibnamefont {Narasgond}}, \bibinfo {author}
  {\bibfnamefont {H.~M.}\ \bibnamefont {Nayfeh}}, \bibinfo {author}
  {\bibfnamefont {E.}~\bibnamefont {Pritchett}}, \bibinfo {author}
  {\bibfnamefont {M.~B.}\ \bibnamefont {Rothwell}}, \bibinfo {author}
  {\bibfnamefont {S.}~\bibnamefont {Srinivasan}}, \bibinfo {author}
  {\bibfnamefont {N.}~\bibnamefont {Sundaresan}}, \bibinfo {author}
  {\bibfnamefont {C.}~\bibnamefont {Wang}}, \bibinfo {author} {\bibfnamefont
  {K.~X.}\ \bibnamefont {Wei}}, \bibinfo {author} {\bibfnamefont {C.~J.}\
  \bibnamefont {Wood}}, \bibinfo {author} {\bibfnamefont {J.-B.}\ \bibnamefont
  {Yau}}, \bibinfo {author} {\bibfnamefont {E.~J.}\ \bibnamefont {Zhang}},
  \bibinfo {author} {\bibfnamefont {O.~E.}\ \bibnamefont {Dial}}, \bibinfo
  {author} {\bibfnamefont {J.~M.}\ \bibnamefont {Chow}},\ and\ \bibinfo
  {author} {\bibfnamefont {J.~M.}\ \bibnamefont {Gambetta}},\ }\bibfield
  {title} {\bibinfo {title} {Demonstration of quantum volume 64 on a
  superconducting quantum computing system},\ }\href
  {https://doi.org/10.1088/2058-9565/abe519} {\bibfield  {journal} {\bibinfo
  {journal} {Quantum Science and Technology}\ }\textbf {\bibinfo {volume}
  {6}},\ \bibinfo {pages} {025020} (\bibinfo {year} {2021})}\BibitemShut
  {NoStop}%
\bibitem [{\citenamefont {Nation}\ \emph {et~al.}(2021)\citenamefont {Nation},
  \citenamefont {Kang}, \citenamefont {Sundaresan},\ and\ \citenamefont
  {Gambetta}}]{nation:2021}%
  \BibitemOpen
  \bibfield  {author} {\bibinfo {author} {\bibfnamefont {P.~D.}\ \bibnamefont
  {Nation}}, \bibinfo {author} {\bibfnamefont {H.}~\bibnamefont {Kang}},
  \bibinfo {author} {\bibfnamefont {N.}~\bibnamefont {Sundaresan}},\ and\
  \bibinfo {author} {\bibfnamefont {J.~M.}\ \bibnamefont {Gambetta}},\
  }\bibfield  {title} {\bibinfo {title} {{S}calable {M}itigation of
  {M}easurement {E}rrors on {Q}uantum {C}omputers},\ }\href
  {https://doi.org/10.1103/PRXQuantum.2.040326} {\bibfield  {journal} {\bibinfo
   {journal} {PRX Quantum}\ }\textbf {\bibinfo {volume} {2}},\ \bibinfo {pages}
  {040326} (\bibinfo {year} {2021})}\BibitemShut {NoStop}%
\bibitem [{\citenamefont {Hunter}(2007)}]{matplotlib}%
  \BibitemOpen
  \bibfield  {author} {\bibinfo {author} {\bibfnamefont {J.~D.}\ \bibnamefont
  {Hunter}},\ }\bibfield  {title} {\bibinfo {title} {{M}atplotlib: A 2{D}
  graphics environment},\ }\href {https://doi.org/10.1109/MCSE.2007.55}
  {\bibfield  {journal} {\bibinfo  {journal} {Computing in Science \&
  Engineering}\ }\textbf {\bibinfo {volume} {9}},\ \bibinfo {pages} {90}
  (\bibinfo {year} {2007})}\BibitemShut {NoStop}%
\end{thebibliography}%
\end{document}